\documentclass[11pt,twoside]{book} 
\usepackage{asp2008n}
\usepackage{times}
\usepackage{lscape}
\usepackage{epsf}

\usepackage{graphicx}
\usepackage{amssymb}
\usepackage{longtable}
\usepackage[figuresright]{rotating}
\usepackage{multirow}

\newcommand{\simless}{\mathbin{\lower 3pt\hbox {$\rlap{\raise 5pt\hbox{$\char'074$}}\mathchar"7218$}}}

\mainmatter

\newlength{\deftabcolsep}
\newcommand{\msun}{M{_{\odot}}}
\newcommand{\arcdeg}{^\circ}
\newcommand{\simgreat}{\gtrsim}
\newcommand{\kms}{{\rm km~s}{^{-1}}}
\newcommand\mdot{\dot{M}} 

\begin{document}

\title{The Dispersed Young Population in Orion}   
\author{C\'esar Brice\~no}   
\affil{Centro de Investigaciones de Astronom{\'\i}a (CIDA)\\
Apartado Postal 264\\
M\'erida 5101-A\\
Venezuela}    

\begin{abstract} 
The Orion OB1 Association, at a distance of roughly 400 pc
and spanning over $\rm 200 \> deg^2$ on the sky,
is one of the largest and nearest OB associations.
With a wide range of ages and environmental conditions,
Orion is an ideal laboratory for investigating fundamental questions
related to the birth of stars and planetary systems.
This rich region exhibits all stages of the star formation process,
from very young, embedded clusters, to older, fully exposed young stars;
it also harbors dense clusters and widely spread populations
in vast, low stellar density areas.
This review focuses on the later, namely,
the low-mass ($M_* \la 2 \msun$), pre-main sequence
population spread over wide spatial scales in Orion OB1,
mostly in the off-cloud areas. As ongoing studies yield
more complete censa it becomes clearer that this "distributed"
or non-clustered population, is as numerous as that
located in the molecular clouds;
modern studies of star formation in Orion would be incomplete
if they did not include this widely spread population.
\end{abstract}



\section{Introduction}

It has been recognized for quite some time that OB associations
are the prime sites for star formation in our Galaxy
\citep[see review by ][]{briceno07a}.
Recent evidence shows that stars in these regions form not only
in dense clusters \citep[e.g.][]{gomezlada98,ladalada03},
but also in more loose aggregates, spread in large numbers over
wide areas spanning tens or even hundreds of square degrees on the sky
\citep{briceno05,slesnick06}. Ongoing surveys suggest that this "distributed"
or non-clustered population, could be as numerous as that in
clusters \citep{briceno05,gutermuth06,briceno07a}.
While the earliest stages of stellar evolution
must be probed with infrared and radio
techniques that can peer into the embedded,
very young (ages $\la 1$ Myr) on-cloud
populations, many fundamental questions including lifetimes
of molecular clouds, cluster dispersal, protoplanetary disk
evolution and triggered star formation, can only
be addressed by surveys of the older, off-cloud populations with
ages $\sim 3-10$ Myr.
Modern studies of star formation in nearby stellar nurseries
are incomplete if they do not include this widely spread population.

The Orion OB1 Association \citep{blaauw64},
located well below the Galactic plane
($-11^\circ \ga b \ga -20^\circ$), at a distance of roughly 400 pc
\citep{ges89,briceno05,hernandez05},
and spanning over $\rm 200 \> deg^2$ on the sky,
is one of the two largest and nearest OB associations
(the other one being Scorpius-Centaurus, see the chapter by
Preibisch \& Mamajek).
\cite{blaauw64} quotes 56 massive stars with spectral types
earlier than B2, more than Scorpius-Centaurus and Lacerta OB1,
and only slightly less than Cepheus. He estimates a total mass
for Orion OB1 of $\sim 8\times 10^3 \msun$, though this number is probably
best interpreted as a lower limit, because it
does not include the lower mass stars.
With a wide range of ages and environmental conditions, Orion
exhibits all stages of the star formation process, from very young, embedded
clusters, to older, fully exposed OB associations, as well as both
clustered and distributed populations.
Therefore, this region
is an ideal laboratory for investigating fundamental questions
related to the birth of stars and planetary systems.
Here we focus on the low-mass ($M_* \la 2 \msun$), pre-main sequence
population widely distributed in the Orion OB1 association,
in the general area within $\alpha_{J2000} \sim 5$h to 6h and
$\delta_{J2000}= \sim +6\arcdeg $ to $-6\arcdeg $.
Regions like $\lambda$ Orionis, which is the subject of the chapter
by Mathieu in this volume, are not discussed here,
 nor the population in the Orion
molecular clouds (chapters by Peterson \& Megeath, Allen \& Hillenbrand),
or clusters like the Orion Nebula Cluster (ONC), discussed by Muench et al.,
or $\sigma$ Ori, which is discussed in the chapter by Walter et al.

\section{The Orion OB1 Association: Historical Overview}

\cite{blaauw64,blaauw91} was the first to identify four major subgroups
in Orion, mostly based on the distribution of O and B type stars,
which he named 1a through 1d.
Orion OB1 encompasses a giant molecular cloud system revealed
in large scale $\rm ^{12}CO$ radio maps \citep[e.g.][]{kut77,maddalena86}.
The two main structures are known as the A and B clouds (e.g. see in
this volume Figure 8 in chapter by Bally et al., also chapters by
Allen \& Davies and Peterson \& Megeath).
The northernmost is the B cloud, showing extended $\rm ^{12}CO$ emission spanning
roughly $8 \deg$, from the L1617 cloud \citep{lynds62} at $\delta \sim +5 \deg$ down to
the Horsehead Nebula and NGC 2024.

\begin{figure}[!ht]
\begin{center}
\includegraphics[width=\textwidth]{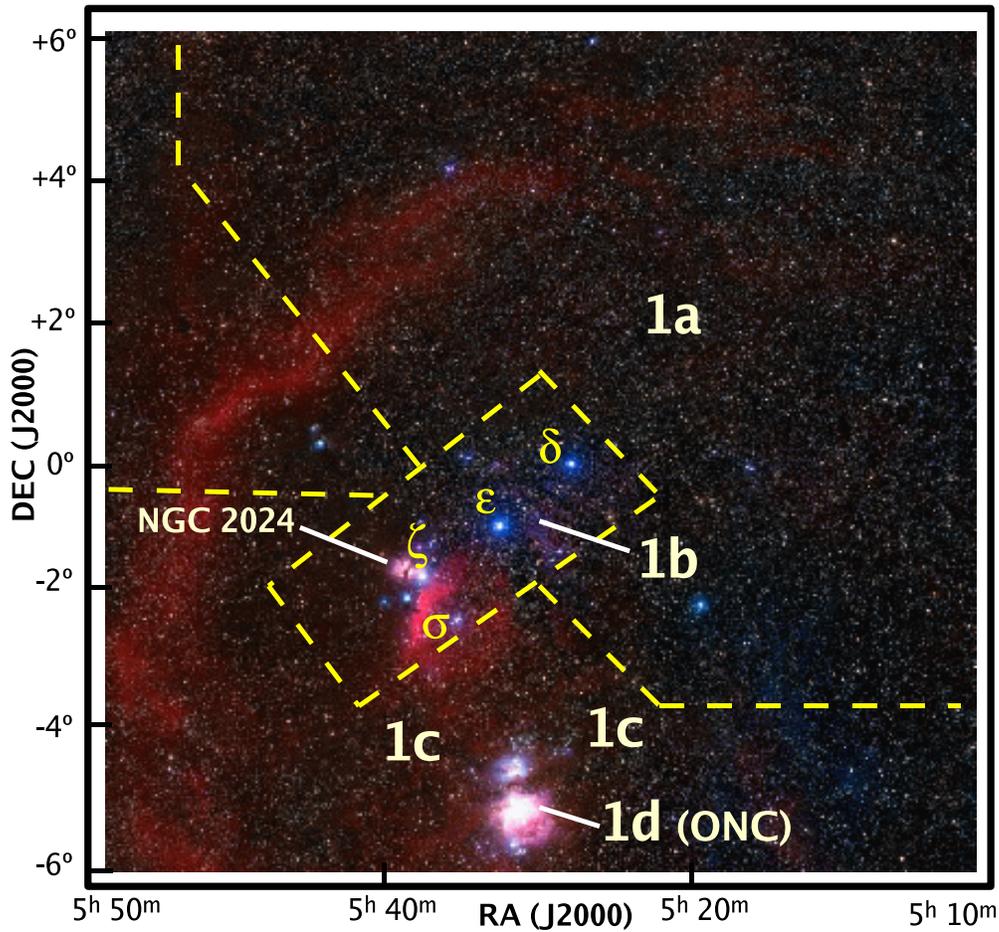}
\caption{\small
Optical wide field image of  $\rm \sim 150 deg^2$ encompassing the
Orion OB1 association. The three Orion belt stars are indicated,
as well as the star $\sigma$ Ori. The dashed lines outline the
boundaries of each sub-association as described in \cite{wh77}.
The older, more extended Ori OB1a region spans the area roughly
N-W of the Orion belt. The Ori OB1b sub-association corresponds
to the area surrounding the three belt stars. The OB1c region
roughly spans the Orion Sword area, around the Orion Nebula,
while Ori OB1d corresponds essentially to the Orion Nebula Cluster.
}
\label{fig1}
\end{center}
\end{figure}

\cite{wh77}
carried out a detailed
photometric study of Blaauw's subassociations,
and were the first to formally assign
boundaries between them (Figure \ref{fig1}).
The Ori OB1a sub-association, encompassing roughly $13\arcdeg \times 8\arcdeg$, a
projected dimension of $\sim 74\times 46$ pc,
extends mostly north and west of the Orion Belt.
Large scale radio \citep[e.g.][]{maddalena86} and
visual extinction maps \citep{schlegel98} show that this wide area is
mostly devoid of gas.
The Ori OB1b region extends over $\sim 4.6\arcdeg \times 2.5\arcdeg$,
a projected dimension of $35 \times 19$ pc, encompassing the three belt stars
$\delta$, $\epsilon$ and $\zeta$ Ori, $\sigma$ Ori with its associated
young cluster, and part of the Orion B molecular cloud, in which are
located the embedded clusters NGC 2023, 2024, NGC 2068 and 2071.
The Ori OB1c group as defined by \cite{blaauw64} includes
the general region of the Orion Sword ($\rm \sim 2 \> deg^2$);
however, \cite{wh77} outline a much larger area for 1c ($\rm \sim 80 \> deg^2$).
The Ori OB1d group is limited to a small area of $\rm \la 0.5 \>deg^2$, encompassing
essentially the Trapezium or ONC.
This general subdivision, based on the early type stellar population
in Orion, was carried further by other studies.
\cite{wh78} divided group 1b into three
subgroups, mainly motivated by \cite{hht64} and \cite{cb66}
who, based on UBV photometric parallaxes,
found a gradient of distances for the
three belt stars, with $\delta$ Ori
being the nearest and $\zeta$ Ori the farthest.
However, more recently \cite{brown94}, using Walraven photometry
of the early type stars in this region,
found no significant differences in distances among the three
belt stars, and no trend in the E-W direction.
Here we will adopt the initial, broader subdivisions proposed by
Blaauw. As we will discuss further on, the issue of subgrouping in Orion
as a whole is probably more complicated than previously thought,
and certainly a problem that cannot be properly investigated by studies
of the early type stars alone.

The spatial distribution of the subgroups as depicted in Figure 4
of \cite{blaauw64} shows an increasing degree of concentration
going from 1a to 1c. If the subgroups are unbound and expanding,
this suggests a sequence of decreasing ages.
In fact, from the color-magnitude diagrams for the hottest stars in each
subassociation, Blaauw provides ages of 12 Myr for 1a, 8 Myr for 1b, 6 Myr
for 1c and $\sim 4$ Myr to 1d (though the kinematic age for the ONC is
listed in his Table II as 0.3 Myr).
\cite{wh77} estimated ages of 7.9, 5.1, 3.7 and  $\la 0.5$ Myr for Ori OB1a
through 1d, respectively.
More recently, \cite{brown94} derived ages for the various subgroups in Orion OB1,
using isochrone fitting in the log $g$-log $T_{eff}$ plane. The resulting ages
were $11.4\pm 1.9$, $1.7\pm 1.1$, and $4.6_{-2.1}^{+1.8}$ Myr for
1a, 1b and 1c respectively, and $< 1$ Myr for group 1d.
The newer results derived from low-mass stars provide better constraints
to the ages of each region, as is shown in Sect. \ref{pmspop}. The trend seen initially,
of Ori OB1a being the oldest sub-association is confirmed, but the actual
age of OB1b is closer to the initial estimates, and significantly older
than the value derived by \cite{brown94}.

The distance to Orion OB1 was quoted by  \cite{blaauw64} as 460 pc. \cite{wh77}
determined distances of 400, 430, 430 and 480 for Ori OB1a, 1b, 1c and 1d
respectively.
\cite{brown94}
used main sequence fitting for the
OB stars in the subassociations
to estimate distances and ages.
They found a difference between 1a and 1b, with 1b having
a similar distance as that of the molecular cloud,
$\sim 440$ pc \citep{grm81}, while
1a is closer, at $\sim 330$ pc.
Stars in \cite{brown94} were
observed by {\sl Hipparcos}; these results
were analyzed by \cite{brown99},
who confirmed the distance difference
between 1a and 1b,
although there were substantial observational errors
in the parallaxes of individual stars.
\cite{hernandez05} used  {\sl Hipparcos} parallaxes
for a subset of stars in Ori OB1a and OB1b, and
found distances of $335\pm 13$ pc for OB1a and
$443 \pm 16$ pc for OB1bc.

Table 1 summarizes our knowledge of the global properties of the
Orion OB1 subgroups, based on studies of early type stars
(spectral types O, B and A).

\setlength{\tabcolsep}{0.7\deftabcolsep}
\begin{table}[!ht]
\caption{Orion OB1 properties derived from high-mass stars}\label{oritable}
\begin{center}
\begin{tabular}{c@{\hskip6pt}c@{\hskip6pt}c@{\hskip6pt}c@{\hskip6pt}c@{\hskip6pt}c@{\hskip6pt}c}
\noalign{\smallskip}
\tableline
\noalign{\smallskip}
Group & Age(Bw64) & Age (WH77) & Age (Bw91) & Age (Br94)   &   D  &  No. Stars  \\
      & (Myr)    &  (Myr)    &   (Myr)   &   (Myr)     & (pc) &   ($M_* > 2 \msun$)    \\
\noalign{\smallskip}
\tableline
\noalign{\smallskip}
  a   &   12      &   7.9      &     12     & $11.4\pm1.9$ & 330  &  234 \\
  b   &   8       &   5.1      &      7     & $1.7\pm 1.1$ & 440  &  123 \\
  c   &   6       &   3.7      &      3?    & $4.6_{-2.1}^{+1.8}$ & 460  &  246 \\
  d   &   4       & $\la 0.5$  &      0     &  $\la 1$     & 480  &   62 \\
\noalign{\smallskip}
\tableline
\noalign{\smallskip}
\multicolumn{7}{l}{\parbox{0.95\textwidth}{\footnotesize
    Note: ages are from \citet{blaauw64,wh77,blaauw91,brown94}.
    Distances and the number of massive stars in each region are
    from \citet{brown94}, except for 1d, for which the number of stars
    with $M_* > 2 \msun$ was estimated from Table 3 of \cite{hill97}.}}\\
\end{tabular}
\end{center}
\end{table}
\setlength{\tabcolsep}{\deftabcolsep}

\section{Wide Area Searches for Low-mass, Young Stars in Orion: \\
Detection Techniques}\label{searches}

Despite being one of the best studied star-forming regions in the solar vicinity,
comprehensive studies of the widely spread, low-mass, pre-main sequence
(PMS) population across Orion OB1 started to become available only quite recently.
The reason has been that, unlike what happens with the very young
on-cloud populations, it is much more difficult to find these older,
more distributed low-mass stars, because the parent molecular clouds
dissipate after a few Myr and no longer serve as markers of
these populations (see also the discussion in Sect. 1
of the chapter by Preibisch and Mamajek).
In order to find these widely spread stars unbiased, wide-field surveys
spanning tens or even hundreds of square degrees are required;
such sensitive searches
have been difficult to carry out in the past.
The majority of studies so far have concentrated on small areas with high stellar density
like the Orion Nebula \citep[e.g.][]{baade37,johnson65,walker69,herbig86,hill97,hill98,herbst00,rebull00,feigelson03,aurora05,preibisch05,stassun06}, $\sigma$ Orionis \citep[e.g.][]{walter97,bejar99,zapatero02,barrado03,sherry04,keny05,jeffries06},
NGC 2024 \citep[e.g.][]{frey95,comeron96,haisch00,haisch01,skinner03,levine06}, NGC 2068
\citep[e.g][]{herbig63,muzerolle05}, and NGC 2071 \citep[e.g.][]{harvey79,dahari99,muzerolle05}
clusters.
Before considering the various search techniques
used to find low-mass PMS stars in the wide off-cloud areas in
OB associations like Orion OB1, it is useful to briefly
review the observational properties of these young low-mass objects.

Low-mass PMS stars, known collectively as T Tauri stars (TTS),
were originally characterized by their late spectral types (G-M),
strong emission lines (especially H$\alpha$), and significant
variability in brightness at most wavelengths; many are also found
spatially associated with regions of dark nebulosity \citep{joy45}.
Stars resembling the original variables first identified as TTS
are currently called "strong emission" or Classical TTS (CTTS).
Subsequent spectroscopic studies of the Ca II H and K lines and the first X-ray
observations with the {\it Einstein} X-ray observatory
\citep{feig89,wku81}
revealed surprisingly strong X-ray activity in TTS,
exceeding the solar levels by several orders of
magnitude, and also revealed a population of X-ray strong objects
lacking the optical signposts of CTTS, like strong
H$\alpha$ emission.  These stars, initially called "naked-T Tauri
stars" \citep{waltmyers86}, are now widely known as "weak-line" TTS (WTTS) after
\cite{herbig88}. At first, the CTTS/WTTS dividing line was set at W(H$\alpha$) $= 10$~\AA.
Recently, \cite{wba03,barradomartin03} revisited the WTTS/CTTS classification
and suggested a modified criterion that takes into account the contrast
effect in H$\alpha$ emission as a function of spectral type in stars cooler
than late K.

\subsection{Objective Prism Searches}\label{objprism}

\begin{figure}[h!]
\centering
\includegraphics[angle=270,scale=0.5]{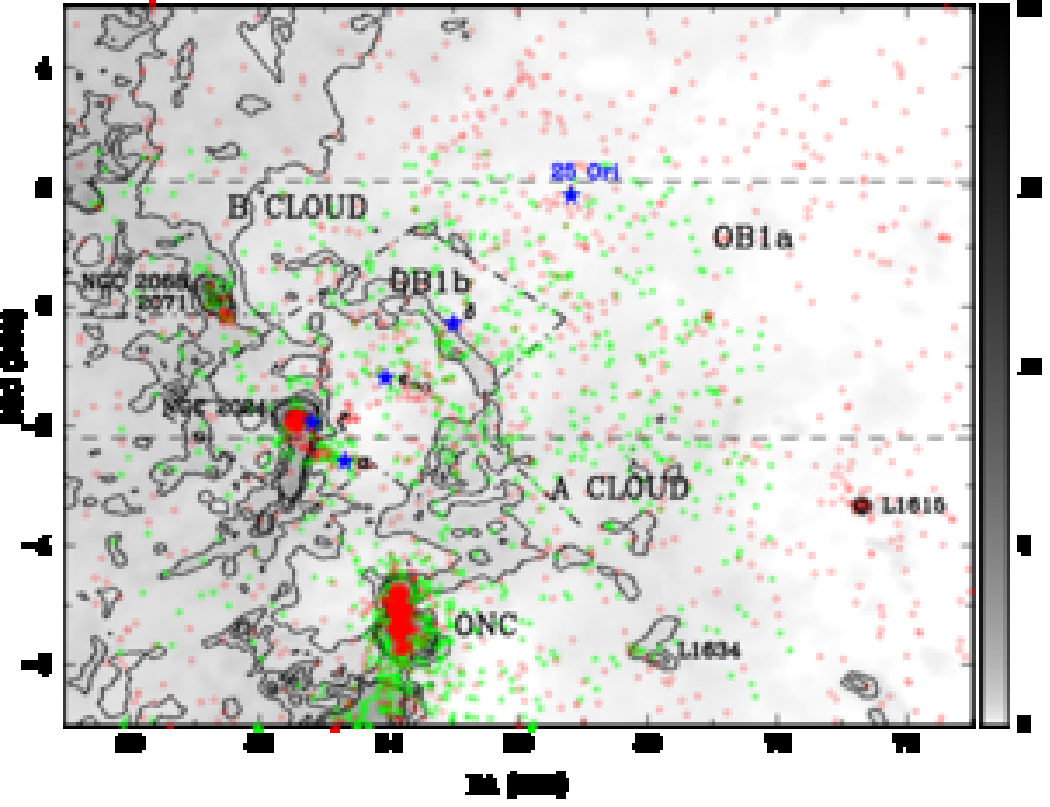}
\caption{\small
Spatial distribution of H$\alpha$ emission and X-ray sources in Orion
listed in the SIMBAD database.
Small red circles are X-ray sources from pointed observations in the
Orion clusters
\citep[e.g., ONC, NGC2024, NGC 2068/2071 ][]{feigelson02,gagne94,getman05,ramirez04,rebull00,tsujimoto02,yamauchi96},
and ROSAT All-Sky Survey sources over the entire OB association \citep{sterzik95,voges99}.
Green starred symbols indicate Kiso H$\alpha$ emission line objects \citep{kyw89,wky89,wky91,wky93}.
The greyscale map shows the integrated  $^{13}$CO emissivity
from \citet{bally87} covering the range from from 0.5 to 20 ${\rm K \, km \, s^{-1}}$.
Isocontours correspond to the extinction map of \cite{schlegel98}.
Dot-dash lines show
the \cite{wh77} boundaries for Ori OB1a and OB1b.
The Orion belt stars, $\sigma$ Ori and 25 Ori are indicated by
large blue starred symbols; also labeled are the
the outlying clouds L1615 and L1634.
}
\label{spatial_rass_kiso}
\end{figure}

The strong H$\alpha$ emission characteristic of CTTS
made large area searches using photographic plates and objective prisms
on wide field instruments like Schmidt telescopes
\citep[e.g.][in Orion]{san71} particularly appealing.
These very low resolution spectroscopic surveys (typical dispersions
of $\sim 1700$~\AA/mm at H$\alpha$)
provided large area coverage, allowed estimates of
spectral types and a qualitative assessment of the strength of
prominent emission lines.
In the Orion OB1 association, the most systematic search was that done with
the 1m Kiso Schmidt \citep{kyw89,wky89,wky91,wky93}
covering roughly 150 square degrees and
detecting $\sim 1200$ emission line stars, many of which were argued to be
likely TTS (Figure \ref{spatial_rass_kiso}).
\cite{weav04} recently identified 63 H$\alpha$ emitting objects
in a deep objective prism survey of the $\sigma$ Orionis region.

\begin{figure}[h]
\centering
\includegraphics[scale=0.4]{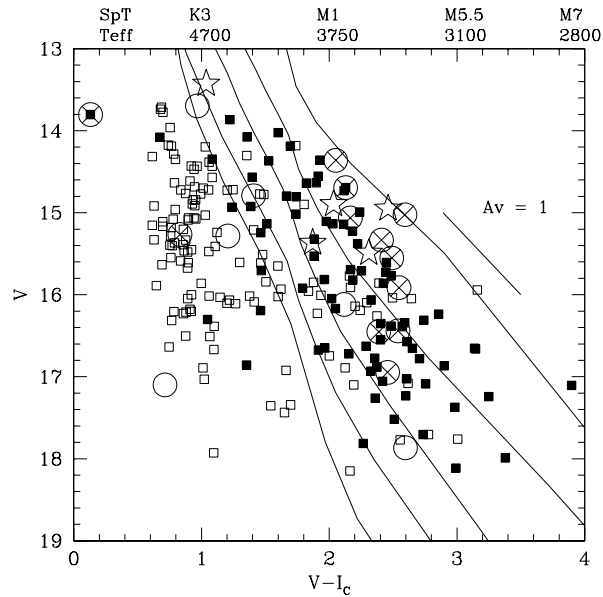}
\vskip -0.2cm
\caption{\small
Color-magnitude diagram for a subset of H$\alpha$-emitting stars
from the Kiso-Schmidt survey, and X-ray sources, in the Orion star
forming region. Kiso stars detected as variables by \cite{briceno05}
are indicated by solid squares; non-variable objects are
plotted as open squares. X-ray sources are plotted as circles,
and those detected as variables by \cite{briceno05} are highlighted by
a $\times$ symbol. The five large stars correspond to previously
known TTS in the area, listed in the \cite{herbig88} catalog.
Solid lines are isochonres from \cite{siess00} at 1, 3, 10, 30, 100 Myr.
}
\label{fig3}
\end{figure}

The main limitation of this technique is the strong bias towards
H$\alpha$-strong PMS stars; few WTTS can be detected at the resolution
of objective prisms \citep[c.f.][]{briceno99}. In fact, out of 25
Kiso sources matching confirmed PMS stars in \cite{briceno05},
only 3 objects (12\%) are WTTS; therefore, specially in somewhat older
regions, in which most TTS are of the WTTS type, objective prism searches
miss nearly 90\% of the PMS stars.
Another caveat is contamination by field stars.
The spatial distribution of the Kiso sources
has been useful to outline the youngest regions in Orion (Figure \ref{spatial_rass_kiso})
where the highest concentrations of CTTS are located \citep{gomezlada98},
but these samples can be dominated by objects like dMe stars in regions far from
the molecular clouds, in which the CTTS/WTTS fraction is small.
\cite{briceno05}
find a CTTS/WTTS fraction of 11\% in Ori OB1a, and a slightly higher value
of 23\% in Ori OB1b,
indicating that the Kiso searches missed the majority of the young
population in these regions.
Only about 50\% of the 218 Kiso H$\alpha$ sources located within the
$\rm 68\> deg^2$ area surveyed by \cite{briceno05} in the OB1a and OB1b
sub-associations, fall above the Zero Age Main Sequence (ZAMS) in the V vs. $\rm V-I_C$
color-magnitude diagram, and just 33\% of those are found to be variable
(a sign that they may be true TTS, see Sect. \ref{variability}),
as would be expected for TTS (Figure \ref{fig3}); these numbers suggest that
 the overall fraction of
PMS objects among Kiso objects is probably not much greater than $\sim 30$\%,
and likely less in the extended area off the main Orion clouds.
Therefore, objective prism studies require follow up spectroscopy to confirm membership.

\subsection{X-ray Surveys}\label{xrays}

X-ray observations are a well established tool
to find active stars, like young PMS stars.
For nearby OB associations, which typically cover
areas in the sky much larger than the field of view of X-ray observatories,
deep pointed observations spanning many square degrees
are usually not feasible.
However, large scale shallow surveys have been conducted
with great success.
The ROSAT All Sky Survey provided coverage of the
whole sky
in the $0.1-2.4$~keV soft X-ray band. With a mean limiting
flux of about $2 \times 10^{-13}\rm\,erg\> s^{-1}\,cm^2$,
this survey provided
a spatially complete, flux-limited sample of X-ray sources
that led to the
detection of hundreds of candidate PMS stars across all of Orion
\citep{sterzik95}. Figure \ref{spatial_rass_kiso}
shows the spatial distribution
of all X-ray sources, both in deep, pointed
observations in selected regions like the ONC, and NGC 2024, 2068, 2071
\citep{feigelson02,gagne94,getman05,ramirez04,rebull00,tsujimoto02,yamauchi96}
and the ROSAT All-Sky Survey sources spread over the entire OB association \citep{sterzik95,voges99}.
The majority of the X-ray sources are concentrated toward the best known
clusters inside the Orion A and B clouds: the ONC, NGC 2024, 2068 and 2071;
this is not surprising, since most deep pointings with X-ray observatories
like {\it Einstein}, {\it ROSAT}, and more recently {\it ASCA}, {\it Chandra} and
{\it XMM} have been done in these regions.
The ROSAT All-Sky Survey sources in the off-cloud areas spanning
Ori OB1a and OB1b are distributed much more uniformly, except for
the strong concentration near the early type star 25 Ori and another
overdensity in the L1615 cloud.
The PMS population in L1615 is discussed in the chapter by Alcal\'a et al. in
this book.
\cite{sterzik95} found a density enhancement of
ROSAT All-Sky Survey sources with a
roughly circular extent of $\sim 5\arcdeg$, centered at $\alpha \approx 82^\circ$,
$\delta \approx +3\arcdeg$.
\cite{briceno05} identified a grouping of PMS stars around 25 Ori,
located at $\alpha_{J2000} \sim 80.8\arcdeg $ and $\delta_{J2000} \sim +1.8\arcdeg $
which is characterized as a populous young cluster by \citet[][see Sect. \ref{25ori} below]{briceno07a}.

The ROSAT All-Sky Survey flux limit corresponds to X-ray luminosities
of about $\rm 10^{30}\>erg\> s^{-1}$ at the distance of Orion OB1
($\sim 400$ pc). This is equivalent to V$\sim 15$
(assuming a typical $log L_X/L_{bol} = -3.6$, \citealt{wolk05}),
the magnitude of a few Myr old K7 star
\citep[$M_* \sim 0.8 \msun$;][]{baraffe98,siess00}.
This implies that the ROSAT All-Sky Survey data
are essentially complete
only for $M \ge 1~\msun$ PMS stars in those regions.
Also, \cite{briceno97} showed
that, because X-ray activity decays slowly during the first 100 Myr
in solar-like stars, the ROSAT All-Sky Survey
candidate PMS samples in areas far from
molecular clouds suffered from significant contamination
by young main sequence stars of ages up to $\sim 10^8 \> yr$.
These limitations have to be kept in mind when working with X-ray
selected samples; at any rate, follow-up observations are necessary
to determine the nature of the objects.

\subsection{Single-epoch Optical Photometry}\label{singlepoch}

Single epoch optical photometric surveys are
frequently used to select candidate
low-mass members of young clusters or associations.
Candidates are usually selected by their
location in the optical color-magnitude diagram (CMD)
above the ZAMS.
The locus in which low-mass PMS stars are located
is usually defined by either a known (spectroscopically
confirmed) population of PMS stars, by comparison with
theoretical model isochrones,
or because the PMS population of the association
is clearly visible as a concentration on the CMD (e.g.,
Figure \ref{vvi_var}).
The main advantage of photometric selection is that for a specified amount
of time on any given telescope a region of the sky can be surveyed to
a fainter limit than can be done by a variability survey, and
low-mass candidate members can be selected without regard to their
brightness variations, which could bias selection against
objects with small amplitudes in their light curves.
The disadvantage of this technique is that there
is inevitably some contamination by
foreground field stars and background giants;
in large areas with a lower density of low-mass association members,
such as Orion OB1b or Orion OB1a the
field star contamination can be large enough to make it difficult
to see the PMS locus (see below).

\begin{figure}[!h]
 \centering
 \includegraphics[scale=0.6]{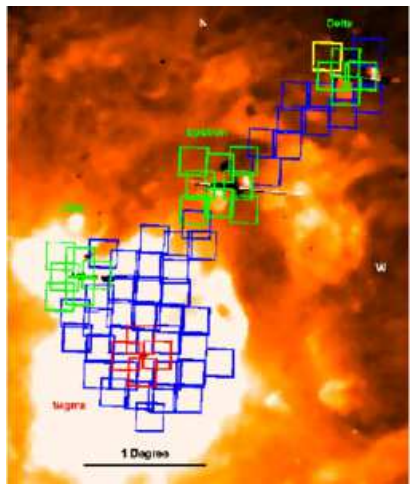}
 \caption{\small
Fields observed by W. Sherry in Ori OB1b. Colors indicate different observing runs. The background image is from the Southern H$\alpha$ All Sky Survey \citep{gaustad01}. The bright regions are the H$\alpha$ emission near $\sigma$ Ori and $\zeta$ Ori. The three belt stars are the dark marks running from left to upper right.}
 \label{sherry_fields}
\end{figure}

The only moderately wide angle, single epoch photometric survey
in the Orion OB1 association has been carried out by \cite{sherry03},
who studied a $\rm \sim 5~deg^2$ area in Ori OB1b and a
small part of Ori OB1a, mostly around
the belt stars (Figure \ref{sherry_fields}).
A subset of this study, centered on $\sigma$ Ori,
was reported by \cite{sherry04}.
Most of the fields shown in Figure \ref{sherry_fields} were observed
in the BVRI filters with the 0.9m telescope at CTIO,
reaching limiting magnitudes $V\sim 21$ and $I_C \sim 19 - 20$.
The identification of the PMS population
was done using a statistical approach, fitting the PMS locus
in color-magnitude diagrams, no spectroscopic confirmation was done.
In Table 3.4 of his PhD thesis, Sherry provides the number
of candidate PMS stars in each region of his survey, totaling
$\sim 800$ objects in the mass range
$M \sim 0.15 - 1\> \msun$. The stars are spread over a
small portion of the Ori OB1a and OB1b regions, and the
inmediate surroundings of $\sigma$ Ori and $\zeta$ Ori.
Data for individual stars only exist for
the likely PMS members of the $\sigma$ Ori cluster
\citep[see Table 4 of ][]{sherry04}.

\subsection{Variability Surveys}\label{variability}

Optical photometric variability is one of the defining characteristics
of PMS stars \citep{joy45,herbig62}. However, due to
limitations in the size of CCD detectors, past work on
photometric variability of young stars have concentrated on follow up
studies of selected samples, that had been identified as PMS
objects by some other means. Only very recently has the
potential of variability as a technique to pick out young stars
amongst the field population started to be realized \citep{briceno01,lamm04}.

\begin{figure}[h]
\plottwo{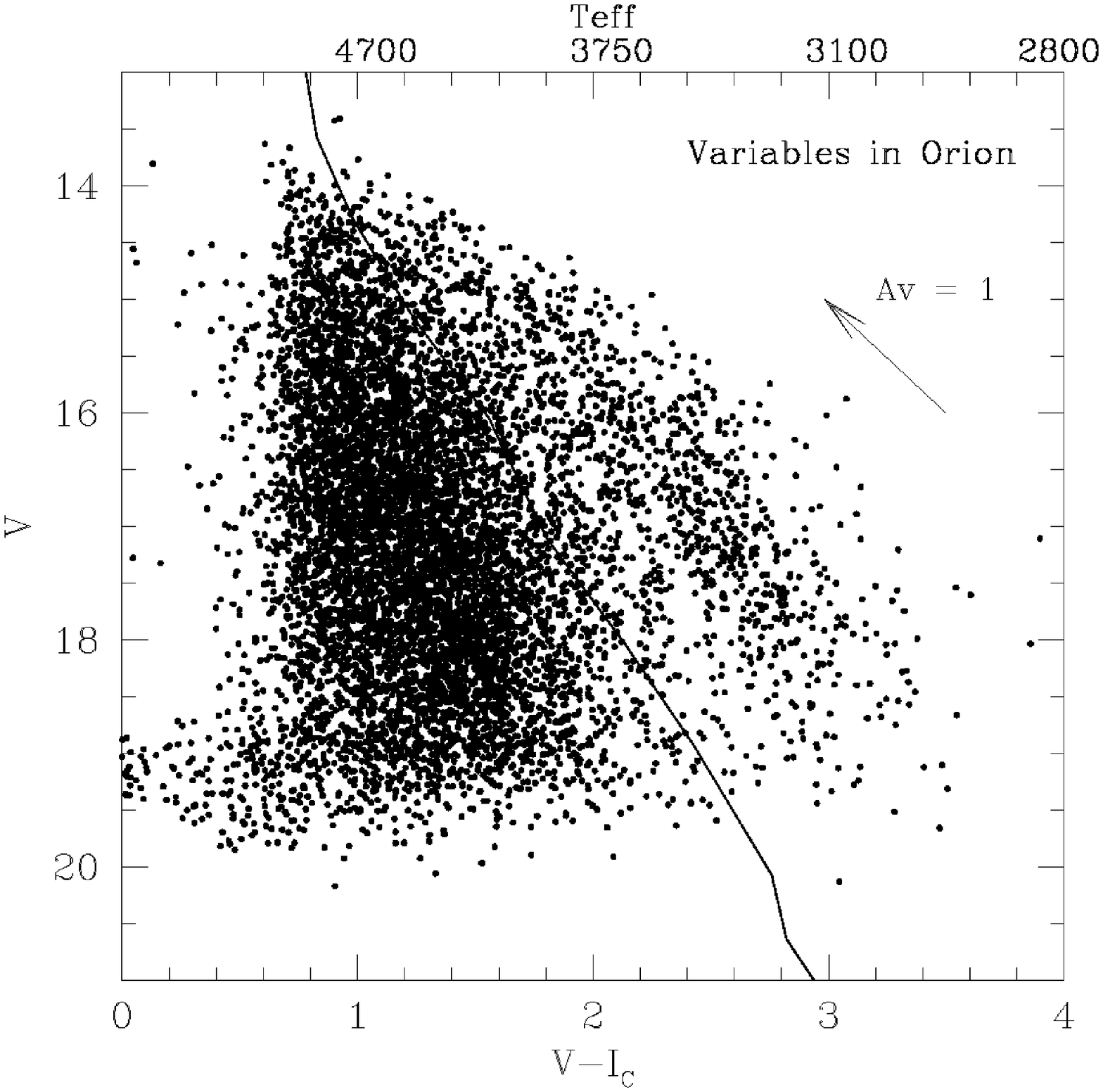}{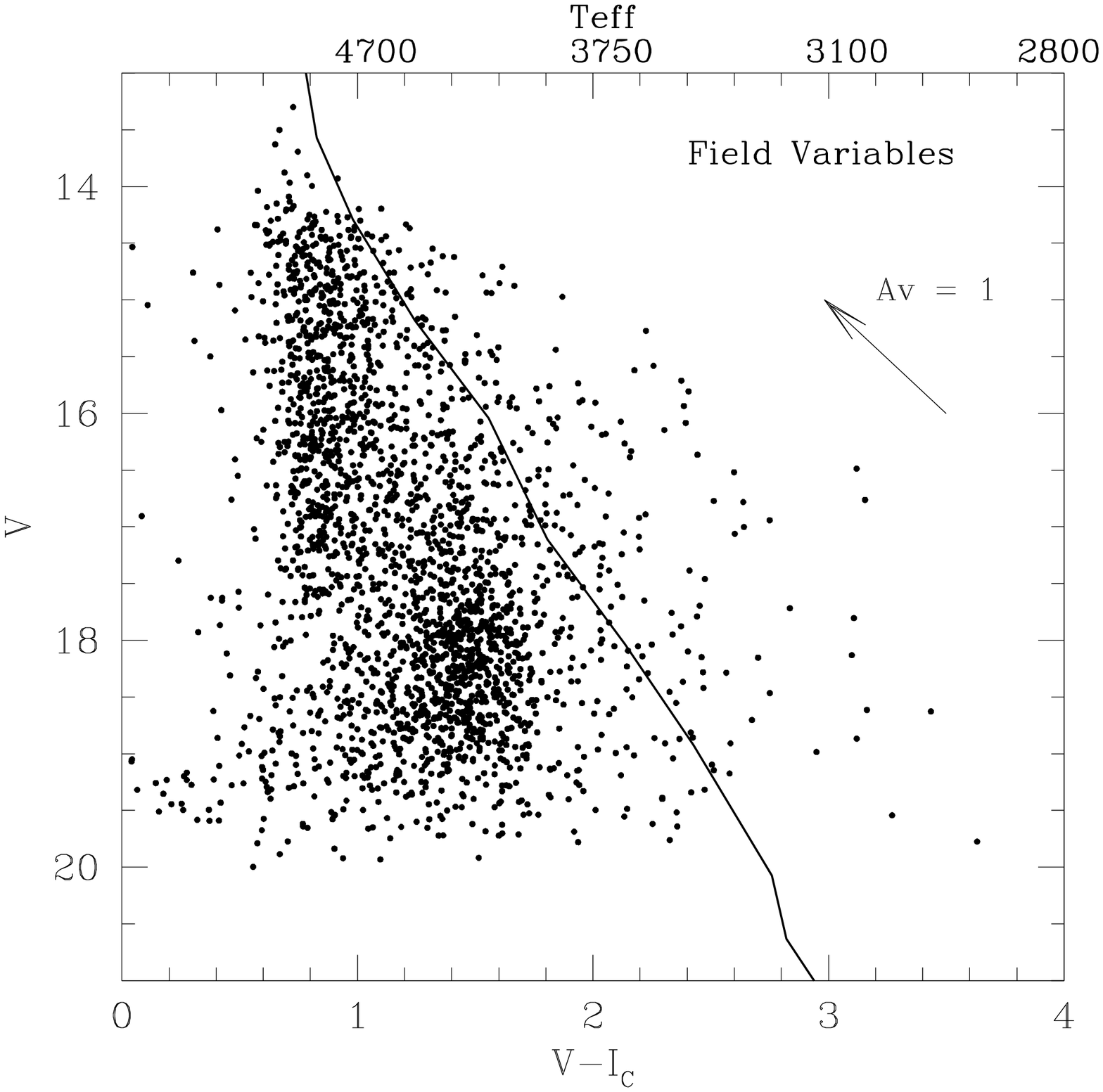}
\caption{\small
Color-magnitude diagrams for stars selected as variable
in \cite{briceno05}.
Left panel: all variables detected in a $2.3\arcdeg $ wide strip
centered at $\delta= -1\arcdeg $ and from $\alpha= 5$h to 6h,
going over part of Orion OB1a and Ori OB1b.
Right panel: control field off Orion; stars in the same strip, with
$\rm \alpha= 4^h48^m - 5^h00^m$.
In both panels the solid line represents the ZAMS
at the distance to Orion \citep[$\rm d\sim 400\> pc$;][]{briceno05}.
The PMS locus is densely populated in the CMD of the Orion OB1 region.
}
\label{vvi_var}
\end{figure}

\cite{briceno01} showed that in surveys spanning wide solid angles,
the large number of objects limits the usefulness of traditional,
single epoch CMDs to single out the PMS locus.
They show in their Figure 2 that
even in a modest region of $\rm \sim 10\> deg^2$ towards Ori OB1b there
can be of the order of 4000 objects above the ZAMS down to V$\sim 19.5$;
but once object lists are filtered by their variability,
only $\la 10$\% remain above the ZAMS.
Moreover, selection of variable stars above the ZAMS clearly
picks a significant fraction of the young members of Orion, as
can be seen in Figure \ref{vvi_var} \citep[from][]{briceno05},
which presents a comparison between CMDs
for a region in OB1b
($\alpha_{J2000} \sim 82.4\arcdeg, \delta_{J2000} \sim 0.3 \arcdeg$)
and a control field
($\alpha_{J2000} \sim 67.5\arcdeg, \delta_{J2000} \sim 0\arcdeg$),
located well off
Orion; the Orion field exhibits an excess of objects above the
ZAMS not present in the control field, even after taking into account
that the Orion region has a higher density of stars
because it is closer to the Galactic plane.
In the Orion field the fraction of variables above the ZAMS is
19\% compared to 10\% in the control field, a factor of $\sim 2$
overdensity. However, the densest
concentration of points in the color-magnitude diagram for the Orion field
occurs at roughly 1 magnitude above the ZAMS. Comparing this
locus in both panels of Figure \ref{vvi_var}, results in
the Orion field having a factor of $3\times$ more objects than the control field.

\begin{figure}[h]
   \centering
	\includegraphics[width=6.5cm]{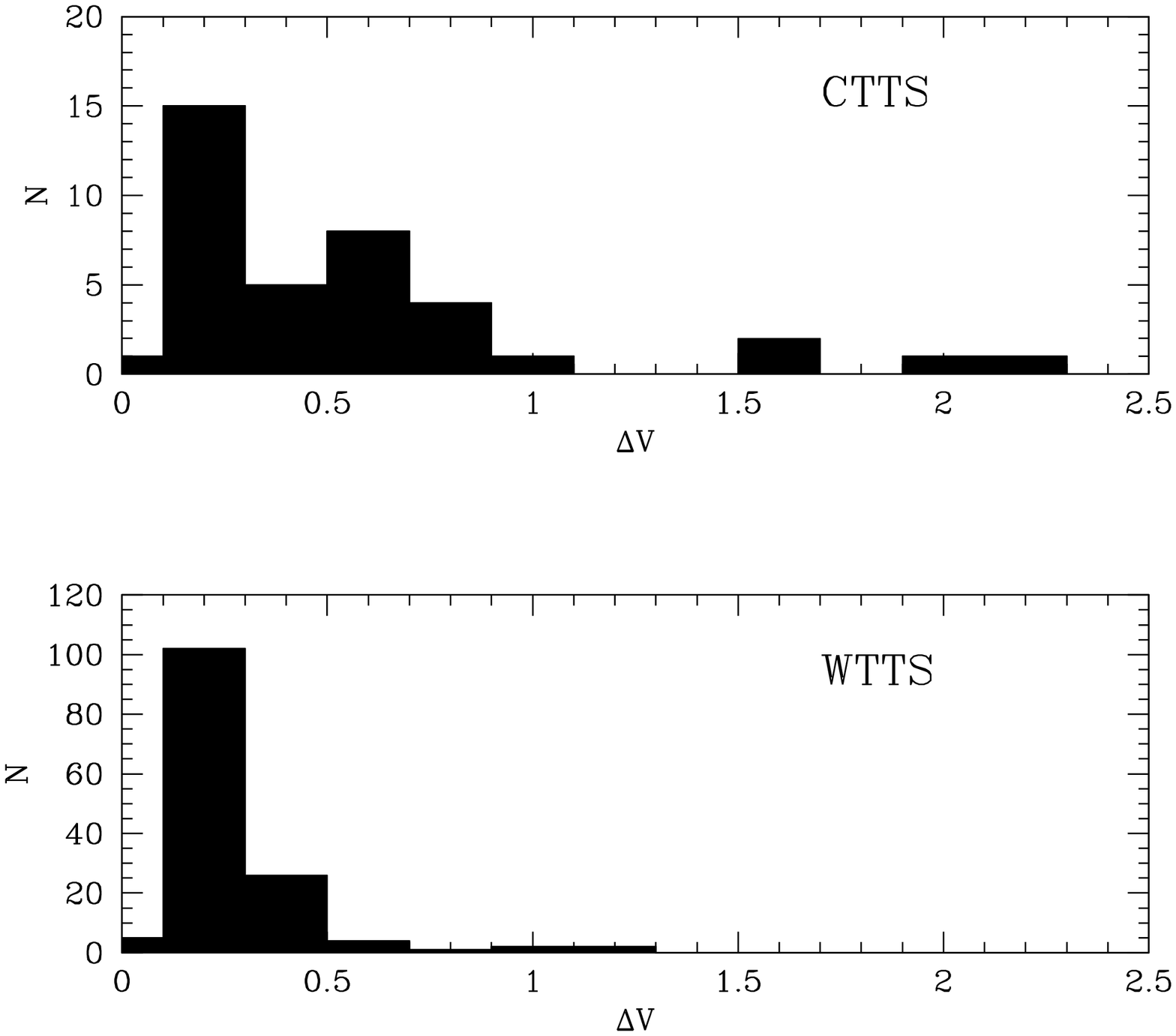}
	\includegraphics[width=5.8cm]{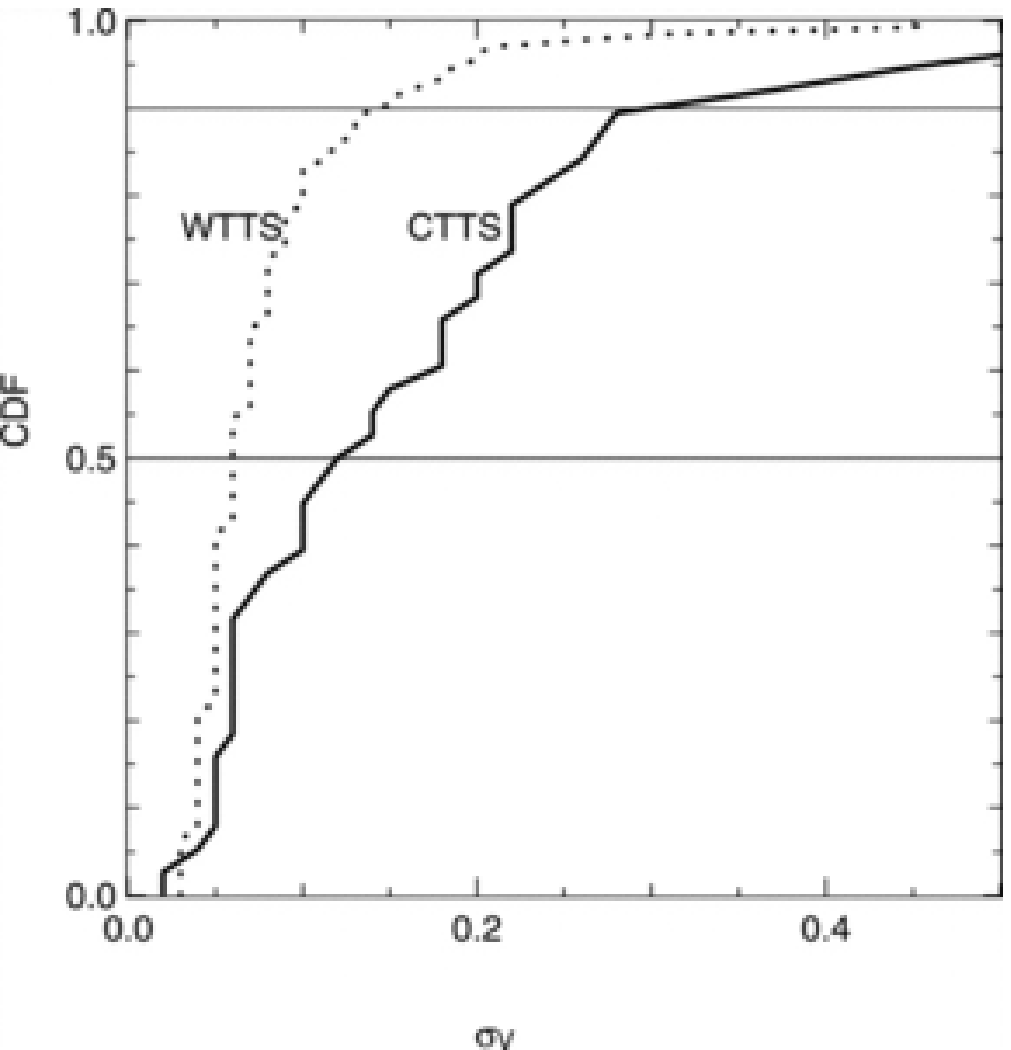}
\vskip -0.2cm
\caption{\small
Left: Histograms showing the distribution of $\Delta {\rm mag}$ for
Classical T Tauri stars (upper panel) and Weak-lined T Tauri stars
(lower panel) in the CVSO \cite[from Fig.~4 of ][]{briceno05}.
Right: Cumulative distribution functions (CDFs)
of $\sigma_V$ for spectroscopically confirmed
WTTSs (dotted line) and CTTSs (solid line) in the sample of \cite{briceno05}.
As can also be seen in the left panel, it is clear that the CTTSs are much
more highly variable than the WTTSs. The median value of $\sigma_V$, at
CDF = 0.5, for the WTTSs and CTTSs is 0.06 and 0.12 mag, respectively.
The 90th percentile values of $\sigma_V$ for the WTTSs and CTTSs have
a greater spread of 0.14 mag \cite[from Fig.6 of ][]{mcgehee06}.
}
\label{sigmav}
\end{figure}

Recently, four major studies have used
large format CCD cameras installed on
wide-field telescopes to conduct multi-epoch,
photometric surveys that use variability to pick out candidate
TTS over extended areas in Orion.
\cite{briceno01,briceno05,briceno07b} have done a VRI variability survey
using the QUEST I CCD Mosaic Camera \citep{baltay02}
installed on the Venezuela 1m Schmidt telescope,
over an area of $\simgreat 150$ square degrees in the Orion OB1
association (CIDA Variability Survey of Orion - CVSO).
The limiting magnitudes are $V\sim 19.7$ and $I_C\sim 19$.
They used a $\chi^2$ test on the
normalized $V$-band magnitudes to select variable stars;
if the probability that the
dispersion is due to random errors
was very low ($\le 0.01$), the object
was flagged as variable.
The minimum $\Delta {\rm mag}$ they detected is
0.06 for $V=15$, 0.09 for $V=17$ and 0.30 for $V=19$.
Figure \ref{sigmav} shows the distribution of V-band amplitudes
for CTTS and WTTS from  \cite{briceno05}.
While both type of stars have a $\Delta V$ peaking at 0.2 mag
the CTTS have a larger spread in amplitudes up to a few magnitudes.

\cite{mcgehee05} analyzed 9 repeated observations over $\rm 25~deg^2$
in Orion, obtained with the Sloan Digital Sky Survey (SDSS).
They selected 507 stars that met their variability criterion in the SDSS g-band.
They did not obtain follow up spectra of their candidates, rather,
they made a statistical analysis to search for photometric
accretion-related signatures.
\cite{mcgehee06} studied a $2.5\arcdeg $ wide strip from
$\rm \alpha_{J2000}=$ 5h to 6h,
encompassing $\rm 37~deg^2$ within the area
of the CVSO survey. He applied reddening invariant
indices to select candidate TTS.
Furthermore, \cite{mcgehee06} use the fact that CTTS are more highly
variable than WTTS  (Figure \ref{sigmav}),
as one of the criteria to distinguish one type of TTS
from the other solely on the basis of the photometric data.
\cite{mcgehee06} used the cumulative distribution function
to set a threshold value of $\sigma_V$
to identify TTS candidates and distinguish WTTS from CTTS.
In general, he identified TTS candidates as those objects with
$\sigma_V \ge 0.05$ mag (in addition to fulfilling other criteria
such as appropriate optical and near-IR colors), and CTTS as objects
above a threshold $\sigma = 0.2$ mag.

Restricted to a much smaller area, \cite{scholz05} carried out a multi-epoch study in the vicinity of $\epsilon$ Ori to explore the rotation and activity of very low-mass PMS objects.
Lacking spectra, they selected 143 objects as young very low-mass stars and brown dwarfs, based on their variability and RIJHK photometry being consistent with membership in this young region. The variability of these objects was investigated using a densely sampled I-band time series covering four consecutive nights with a total of 129 data points per object.

The main disadvantages of variability surveys are that they are
observationaly intensive, not easy to carry out in
highly oversubscribed telescopes, and that, as photometric errors
increase toward fainter magnitudes, the incompleteness is also
larger for lower masses, as more faint objects with small
amplitude variations fail to be detected. However, quantitative
estimates of the actual incompleteness are not yet available.
As with other techniques, spectroscopic confirmation is important.

\subsubsection{Spatial Distribution of Variable Candidate PMS Stars}

The CVSO has revealed a numerous PMS
population in the age range $\sim 4-10$ Myr,
largely discretized in distinct groups of stars \citep{briceno05,briceno07b}.
That stars are born in clusters and
groups is known from studies of molecular clouds \citep{ladalada03},
but that these groups remain as separate entities after the gas is removed
is a new fact, attesting to the rapidness of molecular cloud dissipation.

\begin{figure}[!h]
\centering
    \includegraphics[width=12.0cm]{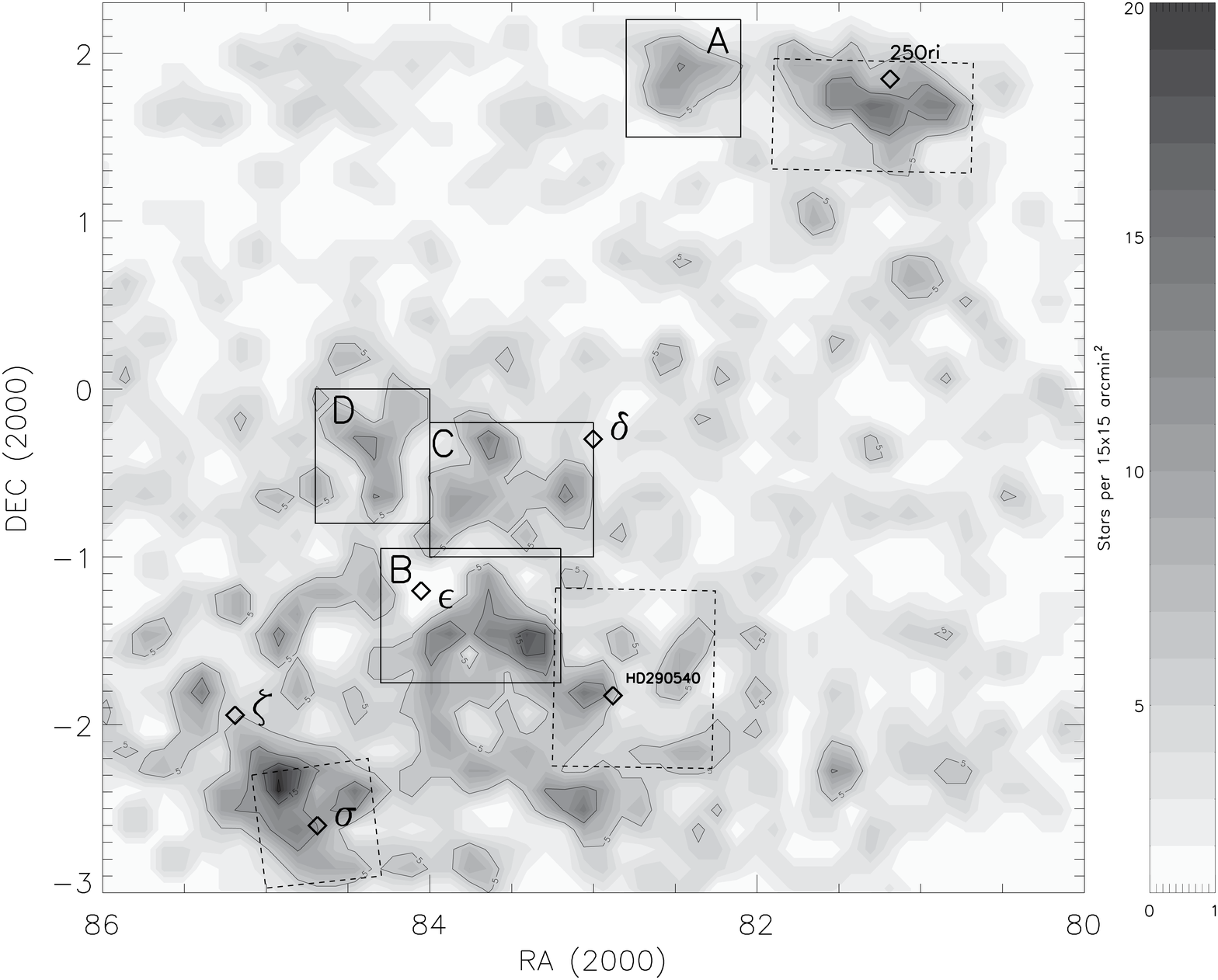}
\vskip -0.5cm
\caption{\small Surface density of 2MASS sources,
located above the ZAMS and flagged as variable,
in Ori OB1a and OB1b \citep{briceno08}. In the upper right
is the 25 Ori cluster \citep{briceno07b}; in the lower left
the $\sigma$ Ori cluster \citep{hernandez07a}.
Dashed-line rectangles
indicate the regions studied by \cite{hernandez07b};
additional fields (solid rectangles) are being observed
with Spitzer by Brice\~no and collaborators
as of the writing of this chapter.
The three stars of the Orion belt are labeled.
}
\label{isodens}
\end{figure}

Figure \ref{isodens} shows the surface density of 2MASS sources,
located above the ZAMS
and flagged as variable, in
the Ori OB 1a and 1b subassociations \citep[data from][]{briceno05,briceno07b}.
The conspicuous overdensity at $\alpha_{J2000}= 81.2^o$, $\delta_{J2000}= +1.85^o$ marks the newly discovered 25 Ori cluster
\citep[``25 Ori''; see Sect. \ref{25ori}][]{briceno07b}.
Other overdensities or "clumps" are also evident in Figure \ref{isodens};
one corresponds to the well known $\sigma$ Ori cluster, with a
peak stellar density similar to that of 25 Ori.

  \cite{mcgehee06} constructed a similar surface density map
to investigate the distribution of candidate TTS picked out
on the basis of $\sigma_g > 0.05$ (see his Figure 8).
He recognized overdensities corresponding to the southern tip
of the 25 Ori group,
the Orion OB1b subassociation, and the NGC 2068/NGC 2071
star formation site in the Lynds 1630 cloud.
The over density that \cite{mcgehee06} finds at the south tip of 25 Ori
corresponds to the two small clumps at $\rm \alpha_{J2000}= 81.0\deg,\delta_{J2000}=+0.7\deg$
and $\rm \alpha_{J2000}= 81.6\deg, \delta_{J2000}=+1.0\deg$ in the map of Figure \ref{isodens},
while the overdensity he sees in Ori OB1b, located at
$\rm \alpha_{J2000}= 83.5\deg,\delta_{J2000}=-0.5\deg$,
corresponds to field C in Figure \ref{isodens}.

These surface density maps demonstrate that variability selection
can trace quite well the young population
in these off-cloud regions.

\subsection{Combination of Single-epoch Optical and Near-IR Photometry}

Single-epoch photometry combining optical and near-IR passbands
has been successfully applied to wide area searches for very low-mass
PMS objects and young brown dwarfs \citep[e.g.][]{briceno02,luh03b}.
In Orion OB1, \cite{bejar03} used I, Z and J-band photometry to search for
very low-mass PMS stars and young brown dwarfs in the vicinity of the Orion belt star $\epsilon$ Ori, spanning an area of $\rm 1000\> arcmin^2$ and reaching a limiting magnitude $I=22$. From the I, I-Z color-magnitude diagram they selected 123 red candidates.

More recently, \cite{downes2008} coadded the existing many epochs of the CVSO
 \citep{briceno05}
to produce single deep R and I-band images, with $3 \sigma$ limiting
magnitudes of $R_{lim} = 21.5$ and $I_{lim} = 20.7$,
and completeness limits of $R_{com} = 20.3$ and $I_{com} = 19.0$,
spanning an area of $\rm \sim 22\> deg^2$ in the Ori OB1a and OB1b
subassociations.
The deep optical imaging, combined
with 2MASS JHK data yield optical/near-IR color-magnitude and
color-color diagrams, that have enabled them to conduct the first
sensitive survey for the lowest mass PMS stars and young brown dwarfs
within the large area observed by the CVSO.

\subsection{Archival Searches: Virtual Observatory}

In a recent work, \cite{caballero08} used Virtual Observatory\footnote{http://www.ivoa.net} tools like Aladin\footnote{http://aladin.u-strasbg.fr/aladin.gml} and TOPCAT\footnote{http://www.star.bris.ac.uk/mbt/topcat/} to search
and combine data from the Tycho-2, DENIS and 2MASS catalogues,
together with photometric, spectroscopic X-ray and astrometric
data from the literature to look for candidate members of the Ori OB1b
subassociation in two 45 arcmin radius fields centered on the belt stars
$\delta$ Ori and $\epsilon$ Ori.
For the bright, early type stars (mostly B and A spectral types) they used optical-near infrared color-magnitude diagrams, the Tycho-2 proper motions, IRAS fluxes and spectroscopic data from the literature to identify likely members. To look for lower mass stars they cross-correlated the DENIS and 2MASS catalogues, and used indicators of youth, like H$\alpha$ or X-ray emission,
and for those cases for which spectra are available, Li~I in absorption or weak Na~I lines indicative of low gravity. In all, they identified 78 bright, early-type stars, and 58 intermediate to low-mass stars, all showing signatures of youth; they also produced a list of 373 photometric candidates.

\section{The Dispersed PMS Population in Orion OB1a and OB1b}\label{pmspop}

\subsection{Low-mass PMS Stars}

Low-resolution follow up spectra are necessary to provide confirmation of
PMS low-mass stars.
These objects can be reliably identified through optical spectroscopy
by the presence of: (1) H$\alpha$ emission,
strong in CTTS, which exhibit equivalent widths
$\rm W(H\alpha) \ge 10${~\AA } at M0,
with larger values at later spectral types \citep[See][]{wba03,barradomartin03},
and weak (a few {\AA}) in non-accreting stars, or WTTS;
(2) the Li~ I $\lambda 6707${~\AA } line
strongly in absorption in late type stars
\citep[$\sim$ K2 and later; e.g.][]{briceno97}
\footnote{It is difficult to use the Li~I line
to select PMS stars of G and early K spectral types,
because in these objects of somewhat higher masses,
lithium suffers little depletion during the PMS phase;
therefore, they can arrive at the ZAMS with Li~I values similar
to the expected primordial abundance. The reason is that in these
stars the convection zone is much shallower, such that material
from the photosphere is not efficiently mixed at depths were
the temperature is high enough to burn Li.};
(3) a weak Na I doublet at 8183 and 8195~\AA, an indicator of surface gravity.

\begin{figure}[tbp]
\centering

\vspace{-3mm}
\includegraphics[width=0.65\textwidth]{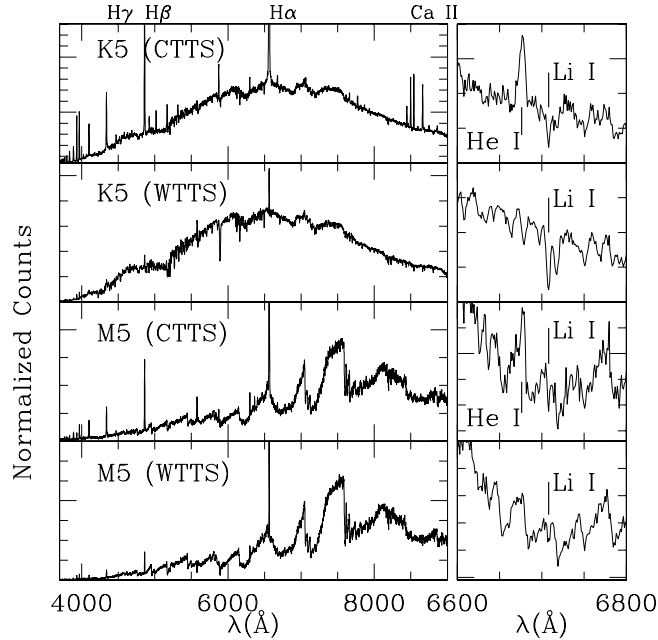}
\vspace{-3mm}
\caption{ \small
  Spectra of members of the Orion OB1
  association, obtained with the multi-fiber spectrograph Hectospec at
  the 6.5m MMT \citep{briceno08}. The upper panels
  show two K5-type PMS objects, an accreting CTTS and a
  non-accreting WTTS.
  For comparison the lower panels show two late spectral type
  members (M5). On the right are the wavelength regions around
  Li~ I $\lambda 6707${~\AA } line for each star. The Li~I line
  is seen clearly in absorption in all objects, next to the Ca I line
  at 6715{~\AA }. In the CTTS the He~I line at 6676{~\AA } is also
  seen in emission.}

\label{spectra1}
\end{figure}

\begin{figure}[tbp]
\centering

\vspace{-3mm}
\includegraphics[width=0.65\textwidth]{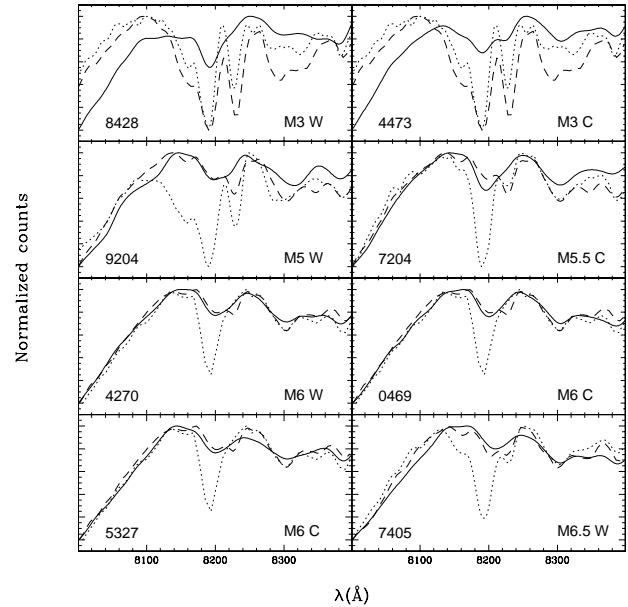}
\vspace{-4mm}
\caption{ \small
  Na I $\lambda 8183, 8195$ absorption lines
  in very low-mass and brown dwarf members of
  Orion OB1 (solid line), compared with
  field dwarfs \citep[dotted line;][]{kir99},
  and with young very low-mass stars and brown dwarfs
  from the Chamaeleon I region
  \citep[$\sim 2$~Myr old, dashed line;][]{luh04a}.
  New members and comparison templates have the same spectral
  types.
  Labels indicate spectral type and class (C=CTTS, W=WTTS).
  Adapted from \cite{downes2008}. }
\label{spectra2}
\end{figure}


Figure \ref{spectra1} shows typical low-resolution ($\sim 6$~\AA)
spectra of Orion PMS members, from K5 to M5 objects, and both CTTS
and WTTS. The M5-type stars exhibit strong TiO bands at
$\lambda \sim 5900, 6200, 6750, 7200, 7700$~\AA,
a feature of cool \citep[$\rm T_{eff}\sim 3200$ K; e.g.][]{kh95},
very low mass \citep[$M\sim 0.15\> M_{\sun}$;][]{baraffe98} PMS stars.
Both CTTS show prominent Balmer lines characteristic of accreting young
stars, and in the case of the K5-type CTTS also the
Ca II IR triplet ($\lambda= 8498, 8542, 8662$~\AA) is seen strongly in
emisssion.
The small panels show a zoom of the region around the Li~ I line, which
is seen well in absorption in all four objects. Contaminating disk field dwarfs
appear similar in all aspects to a WTTS, except in that they lack the
Li~ I absorption line.

 Figure \ref{spectra2} shows the
 Na I absorption lines ($\lambda 8183, 8195$)
in several very low-mass and brown dwarf members of
Orion OB1 \citep{downes2008}, compared with field dwarfs,
and with young very low-mass stars and brown dwarfs
from the $\sim 2$ Myr old Chamaeleon I region.
The weak Na I lines seen in the PMS Orion members
are similar to those of the young Chamaeleon members,
thus providing a strong criterion to select low-gravity young objects,
still contracting toward the ZAMS, and clearly separating them
from late spectral type, foreground dwarfs, which have
much stronger lines.

Here we will consider as low-mass members of the Orion OB1 population
only those stars which have been spectroscopically confirmed
as late spectral type, PMS stars.

\begin{figure}[h!]
\centering
\includegraphics[angle=270,scale=0.55]{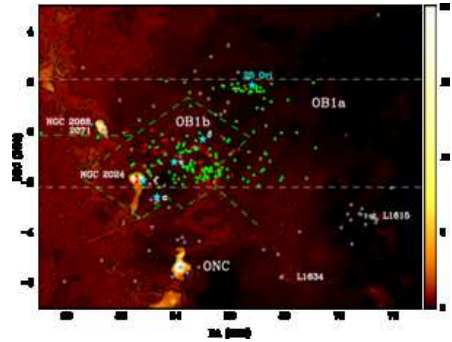}
\caption{\small
Large scale spatial distribution of spectroscopically confirmed, low-mass
PMS stars in the Orion OB1a and OB1b sub-associations, projected
on the dust emission map of \cite{schlegel98}, represented in
false color calibrated to $E(B-V)$ reddening in magnitudes. Contours
are for $A_V= 0.5$ mag.
Stars from \cite{briceno05} are indicated by green dots (WTTS), and
red dots (CTTS).
Small grey stars indicate PMS stars in the \cite{herbig88} catalog.
For clarity, the additional members of 25 Ori and a field in OB1b
from \cite{briceno07b} are not plotted here.
The horizontal dashed lines indicate the limits of the \cite{briceno05} survey.
The dot-dashed lines show the boundaries of Ori OB1a and OB1b
from \cite{wh77}. The three belt stars and 25 Ori are plotted as large
cyan starred symbols.
Stars in the L1615 and L1634 clouds are not discussed here (see chapter by Alcal\'a et al.).
}
\label{spatial}
\end{figure}

There are only a handful of major studies that have undertaken
spectroscopic membership confirmation of low-mass PMS candidates,
across the extended Orion OB1a and OB1b sub-associations.
Figure \ref{spatial} shows the spatial distribution of PMS stars from these
various works across the Orion OB1 subassociation \citep{herbig88,alcala96,briceno05,briceno07b}.
To provide a complete
view of objects known so far, we include in this figure the PMS objects
identified in the L1615, L1616 and L1634 clouds; however, we do not discuss
these regions further as they are the subject of the chapter by Alcal\'a et al.
in this volume.
Also, though a few objects are shown in the Orion B cloud, in and around
NGC 2024, 2068 and 2071 (all in the dark cloud L1630), we will not focus
here on these PMS populations since they are considered in the chapter by Gibb.
Objects in the general ONC area are included in detail in the chapters
by Peterson \& Megeath, and by Muench et al.

Among the first TTS to be confirmed outside the ONC, are those from
the  \textit{Second catalog of emission-line stars of the Orion population}
\citep{herbig72},
mostly from the Lick Observatory H$\alpha$ survey (identified with
the acronym LkH$\alpha$), which later were
incorporated into the Herbig \& Bell Catalog \cite[HBC objects;][]{herbig88}.
\cite{downes88} and \cite{maheswar03} obtained optical spectroscopy of
a number of H$\alpha$ emitting stars, brighter than $V\sim 13$,
identified by \cite{stephenson86}
through objective prism spectroscopy over most of the sky above $b= 10\arcdeg$.

\cite{alcala96} obtained follow-up spectroscopy, and photometry,
of a sample of 181 candidate PMS stars from among 820 ROSAT All-Sky Survey
sources detected
over $\rm 450~deg^2$ in Orion (see their Figure 4).
They identified 112 TTS with spectral types later than F0, the
majority of them WTTS; 54 of these objects are located in the
$\rm 180~deg^2$ area considered here,
between $\alpha=5$h to 6h, and $\delta=-6\arcdeg$ to $+6\arcdeg$.
However, this sample is probably
contaminated by X-ray active, field stars,
especially among the earlier spectral types, for which detection
of Li~I alone does not necessarily discriminate PMS objects from
young stars on the ZAMS.
In fact, the  \cite{alcala96} sample is composed
mostly of G and early-K stars:
28\% late F and G stars, 70\% K stars (K0-K4=60\%,
K5-K7=10\%) and 2\% M stars. \cite{briceno97} predict that
$\sim 40$\% of the ROSAT All-Sky Survey
sources in this region will be X-ray active,
ZAMS stars with G spectral types, and the remaining $\sim 55$\%
will be field K stars, with most in the range K0-K4.
All the G stars and most of the K stars in the \cite{alcala96} sample
have H$\alpha$ in absorption, consistent with a significant fraction
of them being young main sequence stars.
There is an excess of late K stars compared to the expected number of ZAMS stars,
which is interpreted as being due to true PMS stars at these spectral types.
In fact, the ROSAT All-Sky Survey
sources are not uniformly distributed, but rather are seen
mostly projected on or near the molecular clouds \citep[See Figure 4 of][]{alcala96}.
The visual and near infrared magnitudes
of many of the late K and M stars indicate that they are true TTS
with ages of the order of a few times $10^6 \> yr$.

In their large scale variability  survey \cite{briceno01,briceno05}
confirmed 197 TTS through low-resolution ($\sim 6$~\AA) spectroscopy,
with spectral types from mid-K through $\sim $ M4. Comparing with several
evolutionary models they derived ages of 4-6 Myr for the Ori OB1b population
and of 7-10 Myr for Ori OB1a.
The majority of these stars are widely distributed over the
$\rm 68~deg^2$ studied in their initial release.
However, on closer examination, a clear pattern can be seen in the
distribution of CTTS and WTTS (Figure \ref{spatial}). While
WTTS can be found over the entire
region, the CTTS concentrate towards
the regions of higher extinction and
gas density delineating a ring-like structure
with a radius of $\rm \sim 2^\circ$ ($\sim$ 15 pc at
the 400 pc distance to OB1b).
The center of the ring is located
near the B0Iae star $\epsilon$ Ori,
in a region of low $A_V$.
This distribution is similar to
that found in the molecular ring around
$\lambda$ Ori, which has comparable dimensions
to the Ori OB1b ring ($\sim 16 - 20 $ pc,
\citealt[]{dom99,dom01,dom02}).
\cite{dom99}
found that only 5\% of the stars near the
center of the $\lambda$ Ori
ring are CTTS, while most CTTS concentrate near the
molecular ring.
These authors suggested that the
lack of accreting stars near the center
of the region
could be the result of photoevaporation
of disks by the nearby
OB stars. Dolan \& Mathieu proposed
that the CTTS along the ring
were formed by the effects of a
supernova at the center of this structure,
which snow-plowed
pre-existing molecular material into dense
enough concentrations to trigger star formation.
On a smaller scale, a similar situation is also
found by \cite{aurora05}
in a study of the young cluster Trumpler 37,
where the CTTS are strongly concentrated
in a region away from the central O star.
\cite{briceno05} speculate that Ori OB 1b corresponds
to a population formed by a supernova event
which occurred near $\epsilon$ Ori $\sim$ 5 Myr
ago.  Based on their findings, they redefined the boundaries
of OB1b, so that it roughly follows the outer $A_V \sim$  0.5
contour at $\alpha_{J2000} \sim 82.5\arcdeg$ and is limited
by the cloud at $\alpha_{J2000} \sim 85.5\arcdeg$.

In their recent search for Orion OB1 members in two fields
surrounding the belt stars Mintaka ($\delta$ Ori) and Alnilam ($\epsilon$ Ori), \cite{caballero08} report 47 confirmed stars around Mintaka, and 89 in the Alnilam field; for clarity, these sources are not plotted in Figure \ref{spatial}. Among the intermediate to low-mass stars, there are 18 in the Mintaka field and 40 in the Alnilam field.
In the Mintaka field, 2 objects are CVSO stars from \cite{briceno05}, one also in \cite{mcgehee06}, and 10 are Kiso H$\alpha$ emission stars without slit spectra.
In the Alnilam field, 3 objects were reported by \cite{alcala96}, 7 by \citet [][only 3 have split spectra]{bejar03}, 5 by \cite{briceno05}. Among the remaining objects, 15 are H$\alpha$ emission sources from the Kiso objective prism surveys \citep{wky89} and 1 from \cite{stephenson86}, all lacking slit spectra to confirm membership; there are 6 X-ray sources, four of which also have H$\alpha$ emission, though none have slit spectra. In the \cite{herbig88} catalog only PU Ori is within the Alnilam field.  \cite{caballero08} investigated the spatial distribution of the likely members
surrounding Alnilam and Mintaka. They find no compelling evidence for the reality of Collinder 70, also called the ``$\epsilon$ Ori cluster'' \citep[e.g.][]{bejar03}, at least not within their 45 arcmin search radius. Around $\delta$ Ori they find that the radial distribution of stars follows a power-law with an exponent 1-2. They argue that this odd value could be due to contamination from the foreground OB1a population; they call this concentration of likely PMS stars the ``Mintaka cluster'', and propose it may be an evolved version of the $\sigma$ Ori cluster.

The map of the dispersed PMS populations across Orion OB1
is allowing us to start
building a 3-D picture of the giant molecular cloud out
of which the Orion OB1 association formed. Once hundred of pcs in depth,
it formed stars $\sim 10$ Myr  ago in the side closer to
Earth (the nearer OB1a association).
While the gas in this part of the cloud has now dissipated,
star formation is still actively
proceeding on the far side, where molecular clouds A and B remain,
containing a number of young and embedded clusters and many
protostars (see chapters on L1630, L1641, OMC 2/3 in this volume).
In between, a number of groups have been left behind,
the footprint of star formation in this region
(Figures \ref{isodens},\ref{spatial}).
In Table \ref{oritable2} we summarize the properties of the Orion OB1
association as derived from the low-mass stars.

\subsection{Young Brown Dwarfs and Very Low-mass PMS Stars}

So far, only two studies have performed spectroscopic confirmation in
searches for very low-mass PMS stars and young
brown dwarfs in the off-cloud populations of the Orion OB1 association.
In their survey near $\epsilon$ Ori, \cite{bejar03} identified
3 late type objects with H$\alpha$ in emission and Na I line
strengths indicative of low gravity. With spectral types M4.5 to M6,
equivalent to masses in the range $\sim 0.2 - 0.08 \, \msun$ \citep[e.g.][]{baraffe98}, these are all very-low mass PMS stars
and young objects at the substellar limit for this region.

Recently, in their large scale search for PMS objects down to
below the substellar limit, \cite{downes2006,downes2008}
identified one very low-mass star with spectral type M4.5,
one object at the substellar limit, with spectral type M6,
and one brown dwarf with a spectral type M7,
in the Ori OB1a subassociation.
They also found 19 new members in Ori OB1b, 14 very low-mass stars,
out of which 7 have M6 spectral types,
and 5 brown dwarfs (two with spectral types M6.5, two M7 and one M7.5).
Summarizing, they
confirmed 22 new members in the mass range $0.04 - 0.15\, \msun$,
3 in Ori OB1a, one of which is substellar ($M \sim 0.07\, \msun$),
and 19 in Ori OB1b, out of which 7 are at
the substellar limit and 5 are substellar ($M=0.04 - 0.072\, \msun$).

\subsection{The 25 Orionis Cluster}\label{25ori}

\cite{briceno05} identified a clustering of low-mass,
young stars around the early B-type star 25 Orionis (Figure \ref{spatial}).
Through low-resolution follow up spectroscopy of CVSO candidates,
\cite{briceno07b} confirmed nearly 200 PMS stars in this stellar aggregate.

\begin{figure}[h]
\centering
   \includegraphics[scale=0.3]{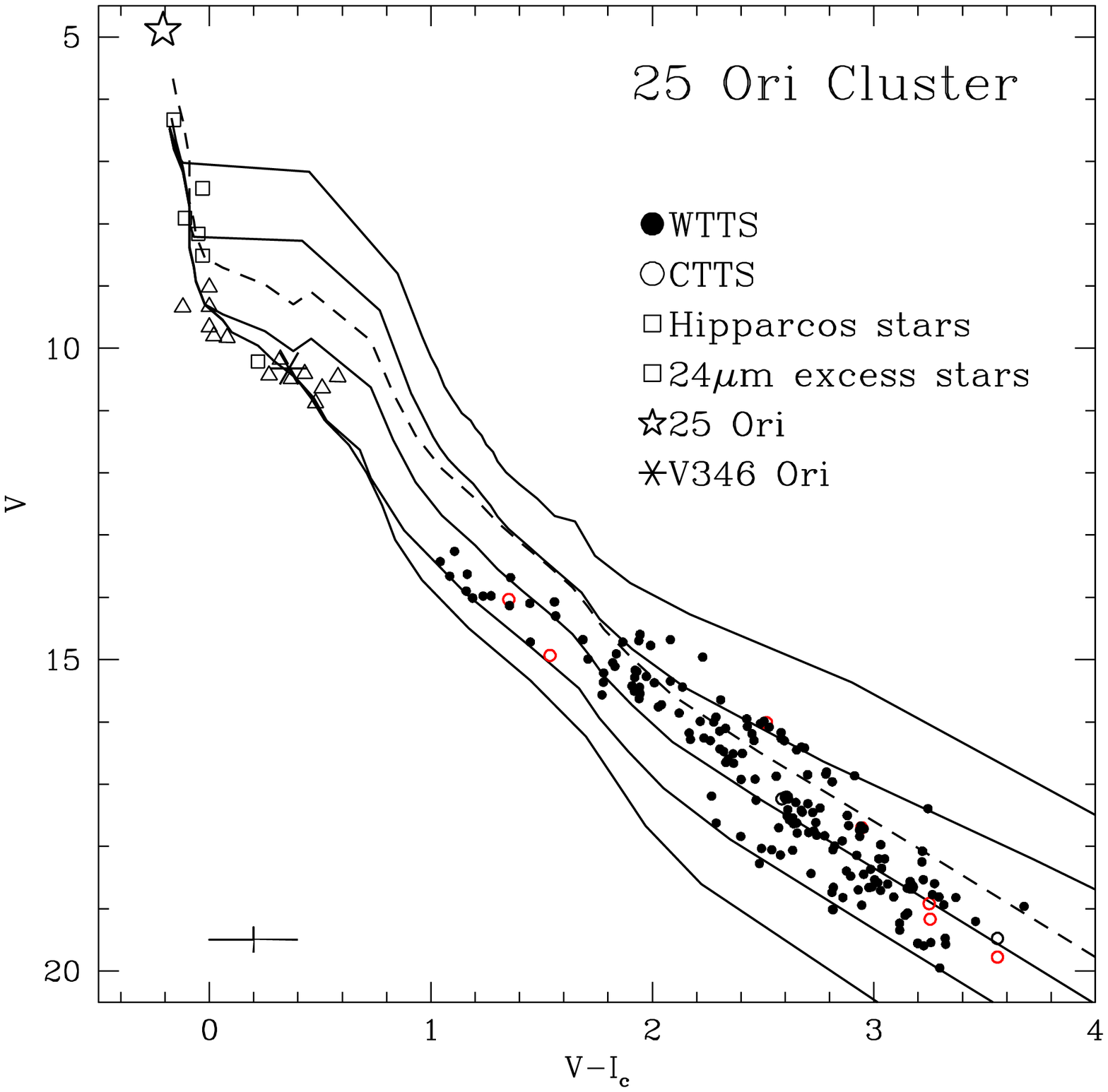}
   \includegraphics[scale=0.3]{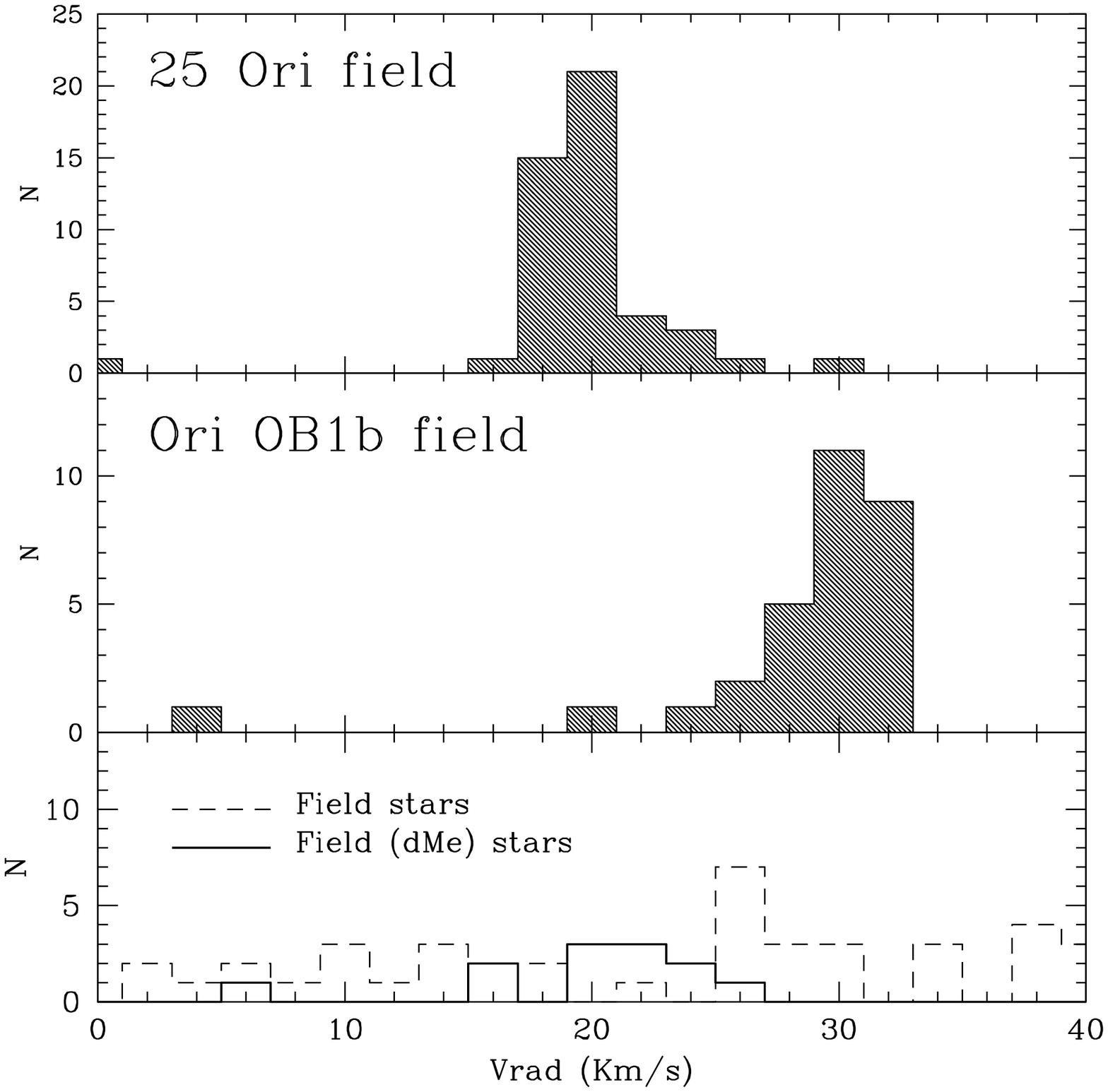}
\vskip -0.5cm
\caption{\small \citep[adapted from][]{briceno07b}.
Left: Color-magnitude diagram of PMS stars in the 25 Ori cluster.
T Tauri stars are indicated with solid circles
(WTTS) and open circles (CTTS).
B type {\sl Hipparcos} stars are shown as open squares.
Open triangles are A-F stars with IR
excesses at 24~$\mu$m \citep{hernandez05}.
Isochrones \citep{siess00} are shown as solid lines,
from top to bottom: 1, 3, 10, 30 and 100 Myr
(which we adopt as the ZAMS).
The dashed line indicates the 0.75 magnitude offset for the
10 Myr isochrone, expected from unresolved binaries.
The error bar at the lower left indicates the
typical uncertainty of the photometry at the faint magnitude limit.
Right: Histogram of heliocentric radial velocities of the 25 Ori cluster
(upper panel) and a field in OB1b with radial velocities
similar to those of the gas in the Orion molecular cloud (middle panel).
The lower panel shows the distribution of radial velocities for field
and dMe stars. Velocity bins are $\rm 2 \> km \, s^{-1}$ wide.
A clear difference
of $\rm \sim 10\>  km \, s^{-1}$ exists between both regions, confirming
the 25 Ori cluster as a distinct kinematic entity.
}
\label{cmd25ori_vrad}
\end{figure}

They found that both the higher mass stars and the TTS in 25 Ori follow
a relatively well defined band in the CMD (Figure \ref{cmd25ori_vrad}).
The spread observed is largely consistent with the upper
limit of 0.75 magnitudes expected from unresolved binaries.
By comparing with different isochrones they derive an age of 7-10 Myr.
Parallaxes for the B and A-type stars in this group indicate that it
is closer than the rest of Ori OB1.
The 25 Ori cluster is the most populous
$\sim 10$ Myr sample yet known within 500 pc.

Many clusters are known in the Orion molecular clouds (e.g. chapters
by Peterson \& Megeath, Allen \& Davis, Muench et al., Walter et al.
in this volume),
but the high degree of spatial substructure among the off-cloud
population, and the discovery of one cluster so far, has been a
surprising new result from recent studies.

\cite{briceno07b} also conducted a radial velocity study of 147 TTS
distributed in two 1 deg wide fields, one located on the 25 Ori group,
and the other in the OB1b region, near the
belt star $\epsilon$ Ori. They found that the 25 Ori members
share a common velocity of $19.7 \kms$, well differentiated from
the $\sim 30 \kms$ velocity that characterizes OB1b members
(Figure \ref{cmd25ori_vrad}), and also
from the $\sim 24 \kms$ velocity that \cite{jeffries06}
assign to the widely distributed,
general population of the OB1a sub-association.
This confirmed that this aggregate is kinematically distinct
from the background molecular cloud, and is probably
a remnant of a now dissipated front end of the Orion giant molecular cloud.

\begin{figure}[h!]
\centering
\includegraphics[scale=0.55]{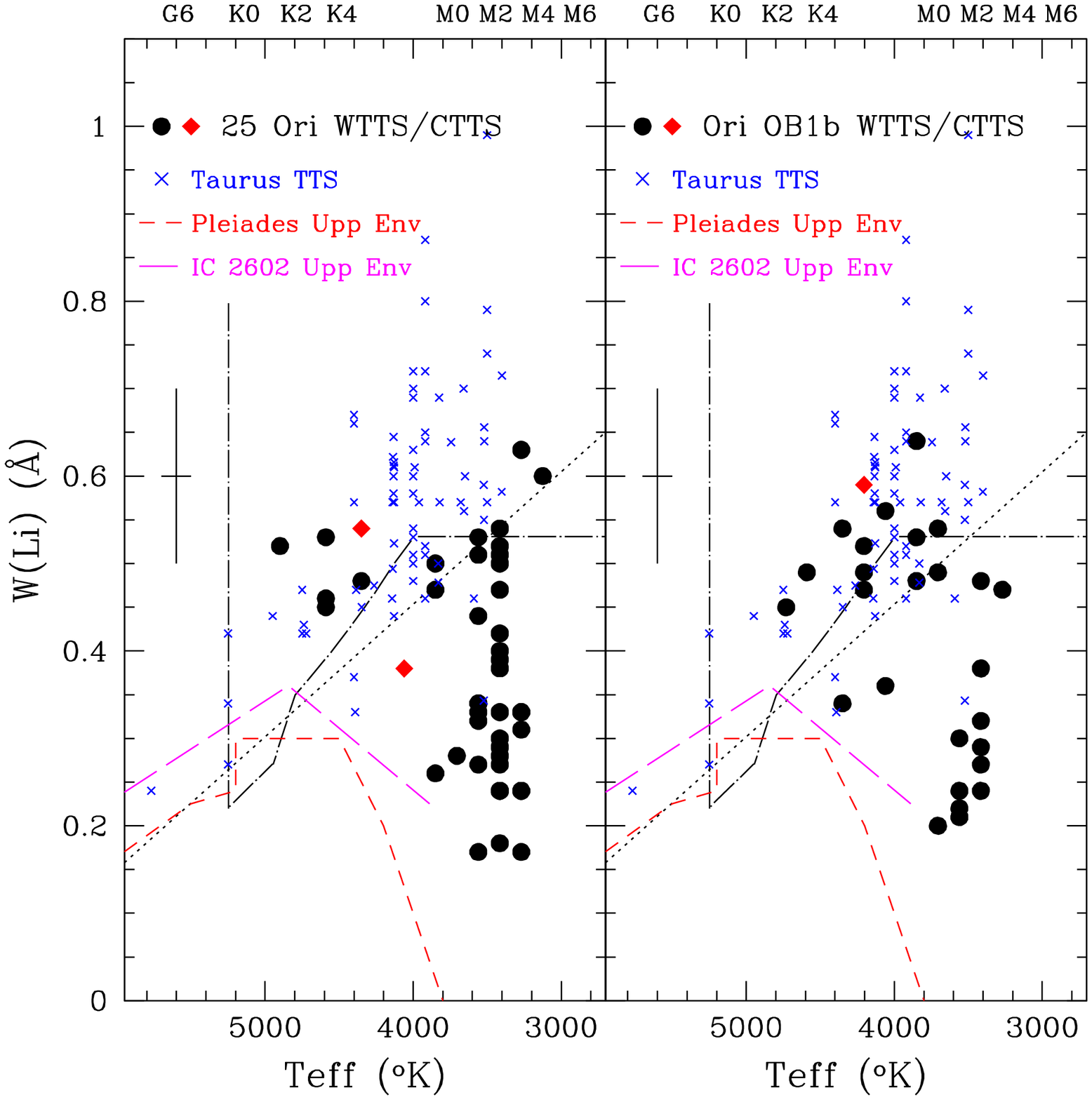}
\vskip -0.5cm
\caption{\small
Equivalent width of the Li~ I 6707{~\AA } line
plotted against the effective temperature \cite[from][]{briceno07b}.
Left panel: stars in the 25 Ori cluster.
Right panel: stars in the Orion OB1b region.
WTTS are shown as dots and CTTS as diamonds.
Taurus PMS stars (age $\sim 1-2$ Myr) are shown as $\times$'s
\citep{basri91,martin94}.
The dotted line traces the lower boundary to the majority of Taurus TTS.
The long dash-dot line corresponds to the lithium isoabundance line
from \cite{martin97}.
The short-dash line is the upper envelope for the Pleiades cluster
\citep[age $\sim 125$ Myr;][]{stauffer98,soderblom93,garcialopez94},
and in the long-dash line the upper envelope of the IC 2602 cluster
\citep[age $\sim 30$ Myr;][]{stauffer97,randich97}.
The typical error bar is also indicated in each panel.\label{lithium}
}
\end{figure}

As mentioned in Sect. \ref{pmspop},
the presence of the Li~ I 6707{~\AA } line strongly in
absorption is a clear indicator of youth in K and M-type stars.
Figure \ref{lithium},
shows the distribution of Li~ I equivalent widths, $\rm W(Li~ I)$,
plotted as a function of the effective temperature for each star
in the sample of \cite{briceno07b}, compared with data for the
$\sim 2$ Myr old Taurus region,  the IC 2602 cluster
\citep[age $\sim 30$ Myr;][]{stauffer97})
and the Pleiades \citep[age $\sim 125$ Myr;][]{stauffer98}.
All the 25 Ori and Ori OB1b members fall above the upper cluster
envelopes; also, in both
Orion regions
the observed $\rm W(Li~ I)_{max} \sim 0.6$ is lower than in Taurus.
Comparison of the $\rm W(Li~ I)$
shows that while 40\% of all the Ori OB1b members
fall within the Taurus TTS locus
only 22\% of the TTS in the 25 Ori group
share this region of the diagram.
The effects of Li depletion among the Orion populations,
seen in the lower $\rm W(Li~ I)_{max}$ and
the increasing fraction of TTS with $\rm W(Li~ I)$ values
below the Taurus lower boundary as a function of age,
are consistent with the ages derived for the Ori OB1b region
and the 25 Ori aggregate, providing further support for
a significant (by a factor $\sim 2$x) age difference between
these two subassociations.

In his study, \cite{mcgehee06}
shows an excess of candidate PMS stars at $\alpha_{J2000} \sim 81\arcdeg$
and $\delta_{J2000}$~=~+0.5$\arcdeg$ to +0.8$\arcdeg$,
which he identifies as the south tip of the 25 Ori cluster (Figure 8
of that article).
Because of this he derives a cluster radius of 8-11 pc,
and suggests  that 25 Ori is an unbound association rather than an open cluster.
However, it is not clear whether the feature identified by McGehee as
the southern tip of the 25 Ori group is physically part of this stellar
aggregate; radial velocities need to be obtained before these
objects can be interpreted as belonging to the 25 Ori cluster.
\cite{briceno07b} presented an initial investigation into the spatial structure
of the 25 Ori group, and argued that the radius of the 25 Ori group is slightly
smaller, $\sim 7$ pc, which combined with the seemingly large number of members
brings into question the actual dynamical state of this stellar aggregate.
\cite{briceno05} found a pronounced peak in the spatial distribution of
low-mass young stars, with a density of $\rm 128\, stars/deg^2$,
at $\alpha_{J2000}=81.3\arcdeg, \delta_{J2000}=+1.5\arcdeg$,
$23.6 \arcmin$ south-east of 25 Ori (see Figure \ref{isodens}).
The density of stars falls off significantly
at a radius of $\sim 1.2\arcdeg$, which corresponds to 7 pc at the assumed
distance of 330 pc. This value is slightly larger than what \cite{sherry04} found for the younger $\sigma$ Ori cluster ($\sim 3-5$ pc).
At a velocity dispersion of $\sim 1 \kms$, an unbound stellar
aggregate would expand roughly 1 pc every 1 Myr.
If the 25 Ori group resembled the $\sigma$ Ori cluster at an
age of $\sim 4$ Myr, a naive dynamical picture would have it
evolve to a cluster radius of $\sim 7-9$ pc
at $\sim 8$ Myr. However, \cite{briceno07b}
caution that because their member census
of the 25 Ori region is not yet complete, especially north of
$\delta_{J2000}=+2.13\arcdeg $, the actual membership and
extent of the group may be larger.

\begin{table}[!h]
\caption{Orion OB1 properties derived from low-mass PMS stars}\label{oritable2}
\begin{center}
\begin{tabular}{ccccc}
\noalign{\smallskip}
\tableline
\noalign{\smallskip}
Group &    Age    &   D  &  No. Stars  & CTTS fraction\\
      &  (Myr)   & (pc) &   ($M_* \la 1 \msun$) & (\%)   \\
\noalign{\smallskip}
\tableline
\noalign{\smallskip}
  OB1a   &   7-10    & 330  &  37$^1$ & 11 \\
  OB1b   &   4-6     & 440  &  142 & 12.6-23 \\
25 Ori & 7-10      & 330  &  197 & 5.6 \\
\noalign{\smallskip}
\tableline
\noalign{\smallskip}

\multicolumn{5}{l}{\parbox{0.8\textwidth}{\footnotesize
    Ages and CTTS fractions are from \citet{briceno05,briceno07b}.
    Distances are from \citet{brown94}.}}\\
\multicolumn{5}{l}{\parbox{0.8\textwidth}{\footnotesize
    $^1$: Number of widely spread PMS stars in OB1a so far, excluding the 25 Ori cluster. }}\\

\end{tabular}
\end{center}
\end{table}

\section{The Initial Mass Function in the Orion Off-cloud Populations}

The form of the
Initial Mass Function (IMF) in OB associations is a crucial information
to determine the number of low-mass stars formed in these regions,
which was not well known until recently, as well as to gain insight
into whether ambient conditions may affect the star formation process.
If the IMF in OB associations is not truncated and
similar to the
field IMF, it would follow that most of their total stellar mass $(\ga 60\%)$
is found in low-mass $(< 2\,M_\odot)$ stars.
This would then imply that most of the current Galactic star formation
is taking place in OB associations \citep{briceno07a}.

Because low-mass PMS stars
($0.7 \ga M/M_{\sun} \ga 0.1$, equivalent to spectral types
later than $\rm \sim K7$) are mostly
on their vertical Hayashi evolutionary tracks,
spectral types roughly correspond to stellar mass,
such that distributions
of spectral types can be used as an approximate proxy of
the IMF in these young regions \citep[ages $\la 10$ Myr; e.g. see ][]{luh03b}.
Moreover, the spectral type is an easily observable
quantity that can be well determined for M-type stars.
Comparisons of these distributions among different
regions have  the advantage of providing a way to look for differences among IMFs without the mediation of theoretical evolutionary models, and the uncertainties and systematics involved in transforming observables such as magnitudes and spectral types to effective temperatures and luminosities.

\begin{figure}[!h]
 \centering
 \includegraphics[scale=0.45]{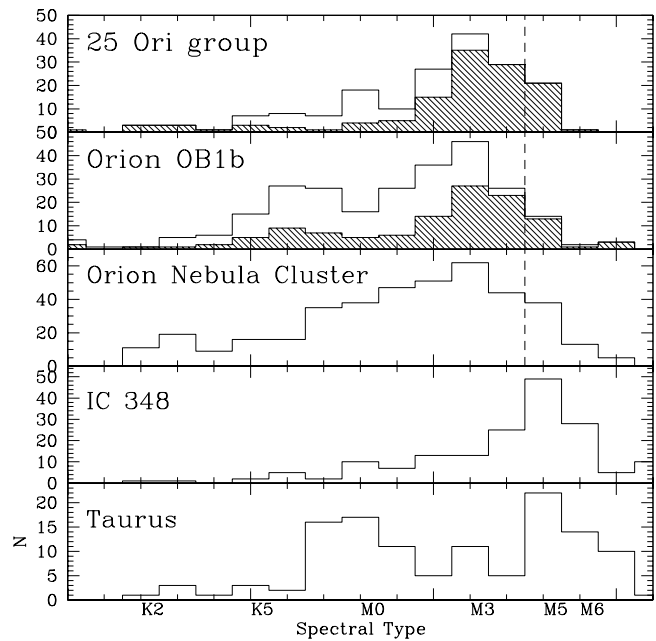}
 \vskip -0.5cm
 \caption{\small Distributions of spectral types for members of
the 25 Ori group and the Orion OB1b subassociation
\citep{briceno05,briceno07b},
compared with the Orion Nebula Cluster
\citep{hill97}, IC 348 \citep{luh03b} and Taurus
\citep{luhman06}.
The filled histograms indicate the kinematic members in 25 Ori and OB1b.
The vertical dashed lines in the Orion panels indicate the approximate completeness
limit. For Taurus and IC 348 the completeness limit is beyond M7.
}
 \label{sptypes_histo}
\end{figure}

Figure \ref{sptypes_histo} shows the spectral type distributions from
mid-K to late M of members of the 25 Ori cluster and the Orion OB1b region
\citep[data from ][]{briceno05,briceno07b}, and compare with similar
histograms for the Orion Nebula Cluster \citep[data from ][]{hill97},
the IC 348 cluster \citep[data from ][]{luh03b} and the Taurus
star forming region \cite[data from ][]{luhman06}.
These samples have all been derived using similar observational
approaches, namely deep photometric searches followed by
spectroscopic confirmation of members. Therefore, strong biases inherent
to differing techniques like X-ray, objective prism surveys or proper
motions, should be minimized.

The distributions for 25 Ori, Ori OB1b and the ONC look remarkably similar,
all peaking at a spectral type M3. This would probably be expected if
these populations do in fact share a common origin in the
same giant molecular cloud complex \citep{briceno07a}.
The histogram for IC 348 has a peak at
a later spectral type of M5, and there seems to be a deficit
of slightly higher mass TTS (spectral types earlier than $\rm \sim M3$.
The distribution of spectral types in Taurus is the one that
seems to differ most respect to Orion; its peak is also located at M5,
as in IC 348, but there is an excess of K7-M0 stars, or
alternatively a deficit of M2-M4 objects.
That the IMF in Taurus seems different to that in other regions has
been noted by various authors \cite[e.g.][]{briceno02,luh03a,luh03b}.
The excess at K7-M0 was
actually the peak of the distribution in earlier studies
\citep[e.g. ][]{briceno02,luh03a,luh04b}, which here is mitigated with the
addition of a number of M5 and M6 new members by \cite{luhman06}.
However, \cite{luhman06}
points out that there is an unknown level of incompleteness
in the spectral type range M2-M6 because of a possible gap
between the faint limits of the wide-field X-ray and
objective prism surveys and the bright limits of deep
optical broadband imaging surveys.
The apparent shift in the peak of the distributions toward
a later spectral type, between Orion and IC 348 and Taurus,
could also be an incompleteness effect; being a factor of $3\times$
further away than Taurus and  $1.5\times$ than IC 348, the
magnitude limited searches in Orion have a larger low-mass limit
(i.e., they are limited at an earlier spectral type).

Except for Taurus,
the distributions in Figure \ref{sptypes_histo} suggest a
rough similarity between the IMF in these regions,
at least within the spectral type range K3-M6.
\cite{luh03b} found that, at least down to spectral
type $\rm \sim M6$, the IC 348 IMF is consistent
with that of the field, a result also obtained for the ONC by
several authors (e.g. Hillenbrand 1997; also see chapter by Muench et al.).
In the  $\sigma$ Ori cluster \cite{sherry04} found that the IMF
in the range $0.1 - 1\, \msun$
is consistent with the \cite{kroupa02} field IMF (see chapter
by Walter et al. in this volume).
In the immediate vicinity of $\lambda$ Ori, \cite{barrado04} derived
an IMF from $0.02 - 1.2~\msun$ that is similar to that derived
in other young regions like the ONC, $\alpha$ Per and the Pleiades.
Over the entire $\lambda$ Ori star-forming region, \cite{dom01}
found that while the global IMF of the $\lambda$ Ori SFR resembles
the field, the IMF on smaller spatial scales
appears to vary substantially
(see chapter by Mathieu for a detailed discussion of the IMF in the
$\lambda$ Ori region).
Summarizing, the variations in the details of the IMF among
differing star-forming regions are still a matter of debate.
As the census of low-mass stars in regions like the Orion OB1
association becomes more complete, we will be in a better position
to assess whether real differences exist, that could be related to the initial
conditions of the star formation process.

\section{Circumstellar Disks and Accretion in Orion OB1a and OB1b}

Circumstellar disks around young stars
play an important role in
determining the final mass of the star and as potential sites for planet
formation. We now recognize that
dust particles suspended in the disk gas evolve,
with solids coagulating and settling toward
the midplane \citep{weid97};
this dust growth and settling are thought to
be the first stage in planetary accumulation \citep{pollack96}.

The presence of disks can be infered by optical signatures
like strong emission in hydrogen Balmer lines like H$\alpha$,
produced by hot (T $\sim 10^4$ K) gas flowing through a
magnetosphere, and by
excess IR emission from warm dust in the disk,
heated by irradiation from the central star.
In order to investigate how these disks evolve and may
give rise to planetary systems,
it is necessary to characterize their properties
in young stellar populations, at ages up to $\sim 10$ Myr.

So far, the most extensive studies of how circumstellar
disk fractions change with
time have been conducted in the Orion OB1 association. However,
the large majority of these investigations have concentrated on the
youngest regions, or densest clusters \citep[see][for a review]{briceno07a}.
Information on disks properties and fractions among the more widely spread
PMS population has become available only very recently.

In the mass range $M \sim 0.3 - 1\, \msun$,
characteristic of low-mass PMS stars,
several studies have started to search for disks among the
dispersed TTS populations.
\cite{briceno05}, using the strong H$\alpha$ emission in CTTS
as a proxy for disk accretion, and applying the CTTS/WTTS classification
scheme by \cite{wba03}, derived the fraction of accreting CTTS across Ori OB1. They obtained an accretor fraction of 11\% in Ori OB1a
and 23\% in OB1b. They compared their estimates with other
regions in their Figure 12.
\cite{briceno07b}, also counting the number of CTTS,
found an accretor fraction of 5.6\% in the  25 Ori cluster,
while in a Ori OB1b field, next to the belt star $\epsilon$ Ori,
the CTTS fraction is 12.6\%.
These values are almost a factor of $\sim 2$ lower than those reported in
\cite{briceno05}. However, they argued that the apparent discrepancy can
be explained because the Ori OB1b sample in \cite{briceno05}
included very young
regions like the area in and around the NGC 2024 cluster ($\la 1$ Myr)
which has a large number of CTTS,
and is nominally located within the \cite{wh77} Ori OB1b boundaries.
Second, as the census of PMS stars in older regions
like Ori OB1a increases, the most frequent type of members
are WTTS, which tends to lower the accretor fraction.

\cite{mcgehee06} combined the SDSS g, r, i, and z bands with 2MASS
JHK magnitudes, to construct reddening-invariant indices and
classify WTTS and CTTS candidates (see Sect. \ref{variability}).
He found a CTTS fraction
of $\sim 10$\%, similar to what \cite{briceno05,briceno07b} derived.
Despite what the absolute CTTS fractions may be in each region,
what does seem like a robust result is the decline
of a factor $\sim 2\times$ in the number of accreting stars between
the $\sim 4$ Myr old OB1b and the 7-10 Myr old OB1a.

\begin{figure}[!h]
\centering
\includegraphics[scale=0.4]{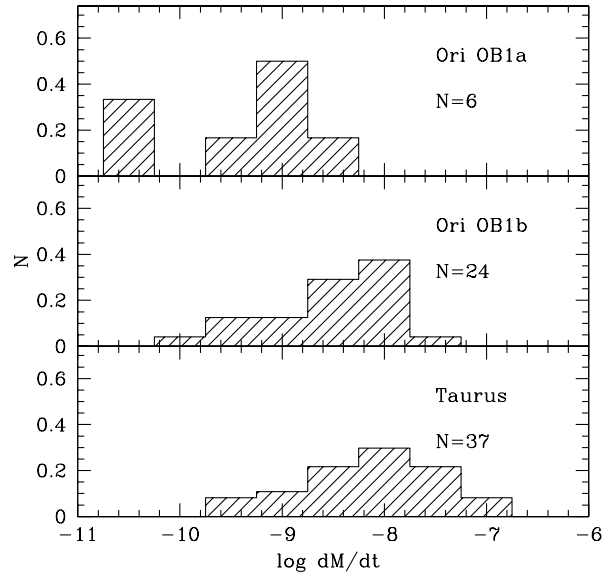}
  \vskip -0.4cm
\caption{\small
Distribution of the disk mass accretion
rates for Ori1 OB1a (upper panel), Ori OB1b (middle
panel) and Taurus (lower panel), from \cite{calvet05a}.
}
\label{fig_distmdot}
\end{figure}

\cite{calvet05a} combined UV, optical, JHKL and 10~$\mu$m
measurements in a sample
of confirmed members of the OB1a and OB1b sub-associations
to study dust emission and disk accretion.
They showed evidence for an overall decrease in IR emission with age,
interpreted as a sign of dust evolution between disks in Ori OB1b
($\sim 4$ Myr), Ori OB1a ($\sim 8$ Myr),
and those of younger populations like Taurus ($\sim 2$ Myr).

Figure \ref{fig_distmdot} shows the distribution of the mass accretion
rates derived by \cite{calvet05a} in Ori OB1a and OB1b,
compared with determinations for the Taurus star-forming region
\citep[mass accretion rates in Taurus are from ][]{gullbring98,hartmann98}.
The age of the three populations increases from bottom to top.
In all three cases, there is
a large spread in the values of $\mdot$.
However, as time proceeds, the number
of rapid accretors decreases, and only low accretors
remain in OB1a.
\cite{calvet05a} applied statistical tests and concluded
that the distributions of accretion rates are significantly
different between the three regions.
This decrease of $\mdot$ with age
qualitatively agrees with expectations from viscous disk evolution.
However, viscous evolution alone cannot explain
the decreasing fraction of accreting objects
with age \citep{muzerolle00,briceno05,briceno07b};
other factors \citep[e.g. inner disk clearing associated
with planet formation,][]{calvet02,dalessio05}
must also play a role in slowing accretion onto the central star.

\begin{figure}[!h]
    \centering
    \includegraphics[width=6.5cm,angle=0]{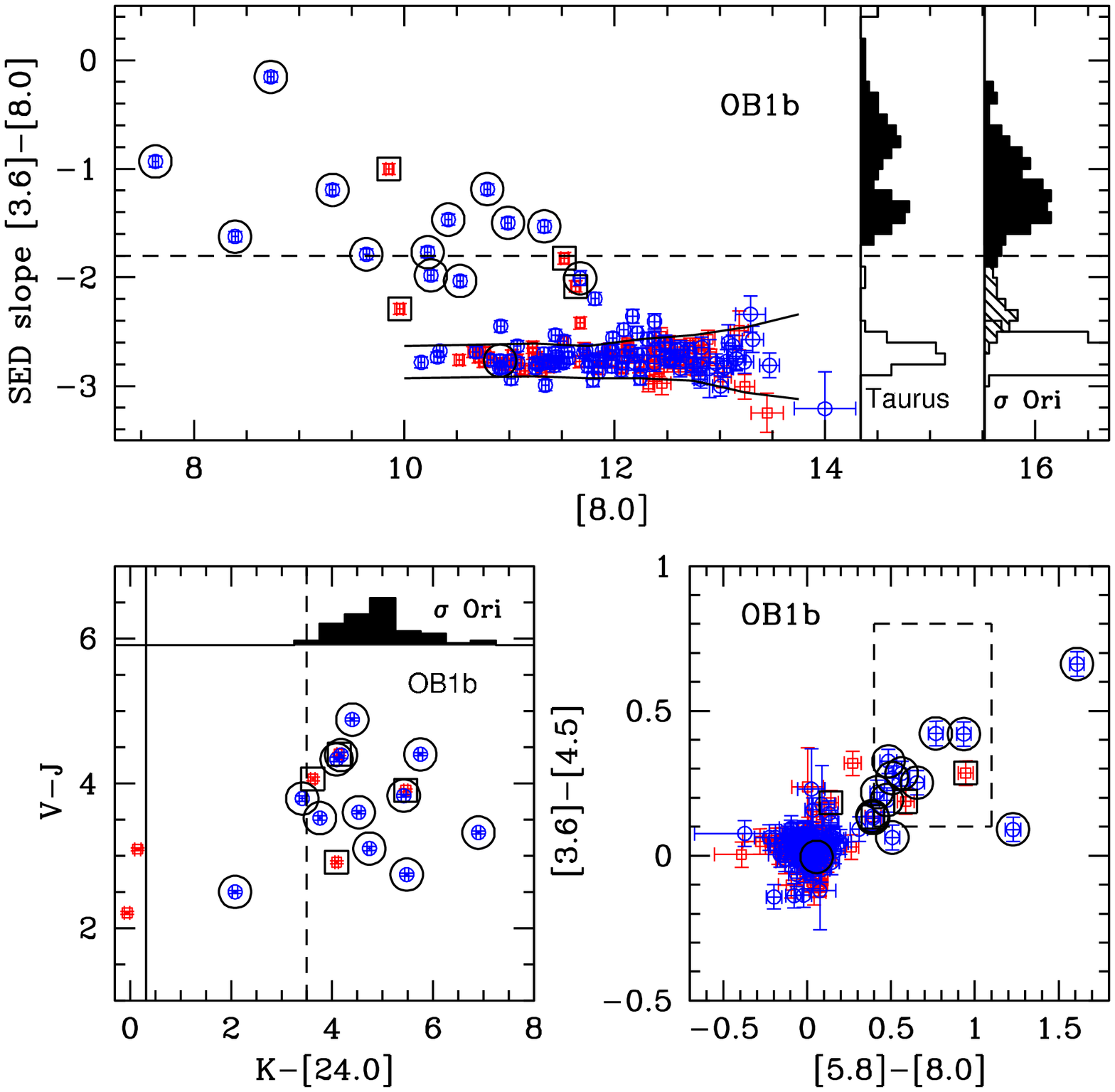}
    \includegraphics[width=6.5cm,angle=0]{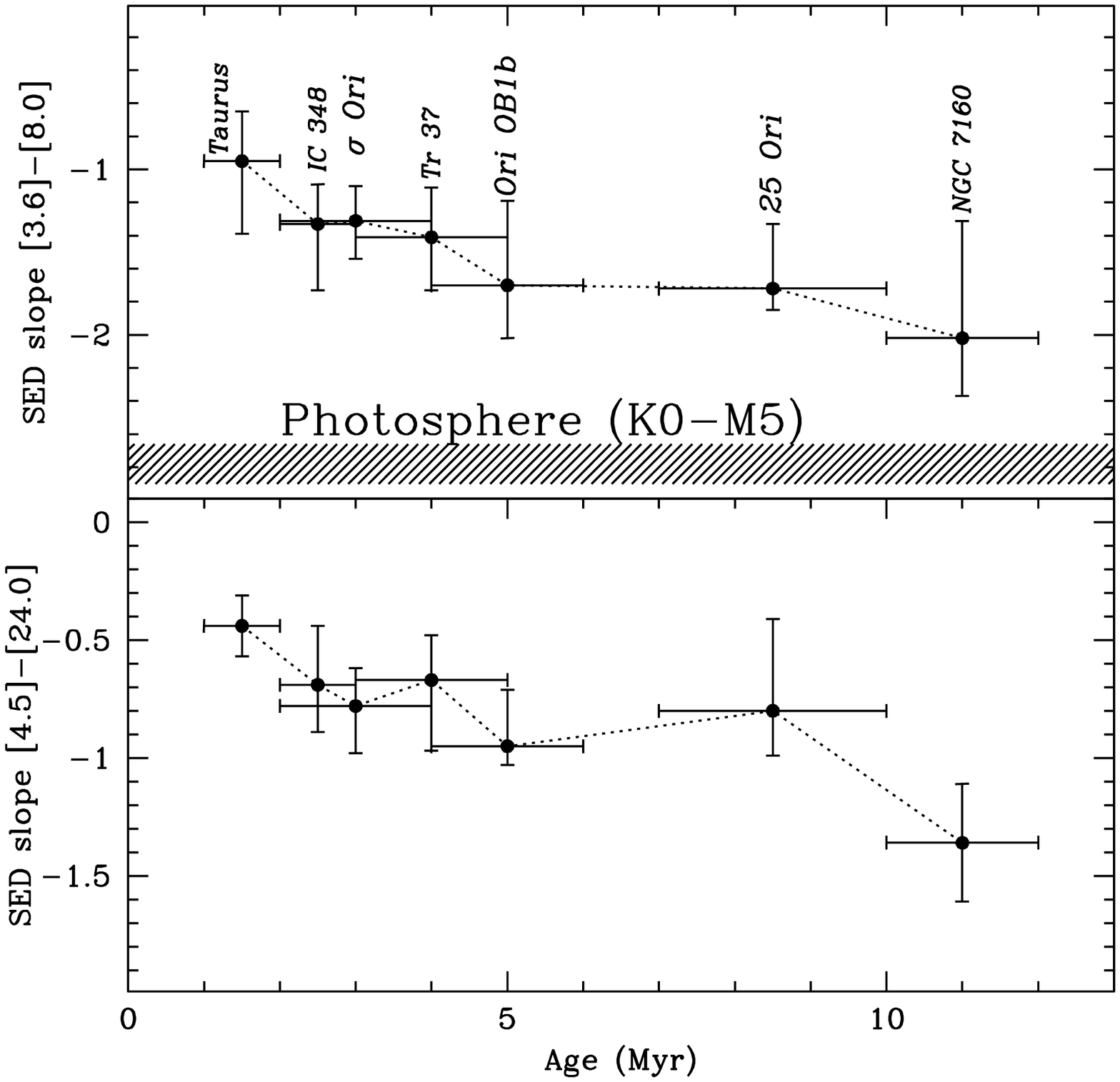}
\vskip -0.2cm
\caption{ \small
Left: Detecting disks with Spitzer \citep[from][]{hernandez07b}.
Members are shown as circles, photometric candidates as squares,
large symbols represent stars with excess at 24~$\mu$m.
Solid histograms indicate IRAC spectral energy distribution
slopes for stars with optically
thick disks in Taurus \citep{hartmann05} and the $\sigma$ Ori cluster
\citep{hernandez07a}; open histograms are stars
with no IRAC excesses. The hatched histogram are objects
with evolved disks in $\sigma$ Ori \citep{hernandez07a}.
Top: IRAC spectral energy distribution
slopes indicating stars with excess at 8~$\mu$m.
The photospheric locus is indicated by the
solid lines. Objects located above the horizontal dashed line are
considered to contain optically thick disks.
Bottom-left: CVSO-2MASS-MIPS color-color diagram indicating stars
with excess emission at 24~$\mu$m. The vertical solid line indicates
the photospheric level, and the
dashed line represents the optically thick disk region limit
\citep{hernandez07a}.
Bottom-right:
IRAC color-color diagram.
The locus of accreting stars is indicated with a dashed line box
\citep{dalessio05}.
Right: Disk evolution.
Upper right: median slope of the spectral energy distributions
between 3.6~$\mu$m and
8~$\mu$m for various stellar groups, including 25 Ori and $\sigma$
Ori. ''Error bars'' are quartiles, i.e., 50\%
of the observations are within these bars.
Lower right: similar plot for the
IRAC 3.6~$\mu$m and MIPS 24~$\mu$m bands.
}
\label{disks}
\end{figure}

More recently, \cite{hernandez07b} used IRAC and MIPS
on Spitzer to look for dusty disks in Ori OB1a and OB1b
(left panel of Figure \ref{disks}).
They found a number of "transition" disk systems,
objects with essentially photospheric fluxes
at wavelengths $\le 4.5 \> \mu$m and excess
emission at longer wavelengths.
These objects can be seen in the left plot of Figure \ref{disks}
(top and lower left panels)
as those symbols falling between the photospheric locus and
the horizontal (or vertical) dashed line that limits the region
were optically thick disks are located.
These systems are interpreted as showing signatures of
inner disk clearing, with optically thin inner regions stretching
out to one or a few AU \citep{calvet02,uchida04,calvet05b,dalessio05,espaillat07}.
They found one transition disk system in 25 Ori and 3 in OB1b,
which represents $\sim 10$\% of the disk-bearing stars, indicating
that the transitional disk phase is short, and therefore hinting at a
rapid shut off of the accretion phase in these systems.

\cite{hernandez07b} derived disk fractions of 6\% in
the 25 Ori aggregate and 13\% in Ori OB1b, similar to what
\cite{briceno07b} found from accretion indicators.
\cite{hernandez07b} also confirmed a decline in IR emission
by the age of Orion OB1b and, by comparing the infrared
excess in the IRAC and MIPS bands among several stellar groups, show
that not only does inner disk emission decay with stellar age,
but the inner disk dissipates more rapidly.
In the upper right panel of Figure \ref{disks} there is a clear
decrease of the median slope of the IRAC
spectral energy distribution
with age, approaching the photospheric limit,
indicating a decrease of optical depth as dust grows and settles.
There is also a decrease in the slope of the spectral energy distribution
in the IRAC-MIPS diagram (lower right panel in Figure \ref{disks}),
but slower than in the IRAC-only plot; the IRAC-MIPS slope
corresponds to regions further out in the disk
than the IRAC-only slope.
The faster decrease in the IR dust emission
from the inner parts of the disk is
indicative of an ``inside-out'' clearing.

In their sample of very low-mass TTS and young brown dwarfs,
\cite{downes2008} looked for the presence of accretion signatures by measuring
the strength of the H$\alpha$ line in emission. Using the
criteria by \cite{wba03} they classified
their new members as having CTTS or WTTS nature.
They found that the 3 new members confirmed in Ori OB1a are WTTSs,
while $39^{+25}_{-22}\%$ of the new members in Ori OB1b exhibit
CTTS-like behavior, suggestive
of ongoing accretion from a circum(sub)stellar disk.
They also found
that none of the members confirmed in OB1a show near-IR color excess
while $38^{+26}_{-21}\%$ of OB1b members show H-K color excess.
These results are consistent with findings by \cite{briceno05}
for higher mass TTS in Orion OB1.
The similarity in CTTS-like properties and near-IR excess
across the substellar boundary gives support to the idea
of a common formation mechanism for low mass stars and
at least the most massive brown dwarfs.

Finally, among the higher mass members of the widely
spread PMS population of Ori OB1 and OB1b
(stars of spectral types B, A and F), so far
only two studies have looked for circumstellar disks.
\cite{hernandez05} carried out a study of the early-type
stars in Orion OB1 (among other nearby OB associations),
in order to identify Herbig Ae/Be (HAeBe) type stars, the PMS
higher mass counterpart of TTS. They studied B, A, and F stars
with membership determined from Hipparcos data.
Comparing the data from associations with different ages,
and assuming that the near-IR excess in the HAeBe stars arises
from optically thick dusty inner disks, they found that
the inner disk frequency in the age range 3-10 Myr in
these higher mass stars is lower than that in the
low-mass stars ($<1 \,\msun$), in particular,
a factor of $\sim 10$ lower at $\sim 3$ Myr.
This indicates that the timescale for disk evolution is
much shorter in young stars with masses in the range
$2\, \msun \la M \la 10\, \msun$, which could
be a consequence of more efficient mechanisms of
inner disk dispersal.

\cite{hernandez06}
compared Spitzer 24~$\mu$m observations of B, A and F stars
in Ori OB1a and OB1b with similar objects in
other stellar groups, spanning a range of ages from 2.5 to 150 Myr,
and found that debris disks are more frequent and have larger
24~$\mu$m excess at ages $\sim 10$ Myr (OB1a).
This trend agrees with predictions of models of evolution
of solids in the outer regions of disks \citep[$> 30$ AU; e.g.][]{kenyonbromley05},
where large icy objects ($\sim 1000$ km) begin to form at
$\sim 10$ Myr; the presence of these objects in the disk
initiates a collisional cascade, producing enough dust particles
to explain the relatively large 24~$\mu$m excesses observed
in Ori OB1a.
Combining Spitzer observations, optical spectra, and 2MASS data,
they also identified a new Herbig Ae/Be star (HD 290543)
and a star (HD 36444) with a large 24~$\mu$m excess, both in OB1b.
This last object can be explained as a intermediate stage
between a Herbig Ae/Be type object (which harbor optically thick
disks) and true debris systems, or alternatively
as a massive debris disk produced by a collision
between two large objects ($> 1000$ km).

\vspace{0.5cm}

{\bf Acknowledgements.}
I am deeply indebted to my close collaborators
Nuria Calvet, Lee Hartmann, Jesus Hern\'andez,
Kathy Vivas, Juan Jose Downes, Lori Allen and James Muzerolle;
without their hard work, deep insight and lively discussions
the PMS map of the Orion OB1 association
would still be largely blank; all the merits of the
CIDA Variability Survey of Orion (CVSO)
and related studies are theirs, the errors are solely mine.

The CVSO has been made possible by the
continuous efforts of the CIDA staff, observers and
Night Assistants, in particular,
O. Contreras, F. Moreno, G. Rojas and
U. Sanchez, technical staff, especially Gerardo S\'anchez, and
through the painstaking work by Perry
Berlind and Mike Calkins at SAO, who have
obtained spectra for thousands of candidate PMS stars in Orion.
I also thank Susan Tokarz at CfA for the reduction of
the spectroscopic material obtained at SAO.
I am grateful to the referee for useful comments,
and very specially to the editor, Bo Reipurth, for his endless
patience and his further input to the manuscript.
The CVSO has received support from grants S1-2001001144 of
FONACIT, Venezuela, NSF grant AST-9987367 and NASA grant NAG5-10545.

Much of the results compiled here are based on observations obtained at the
Llano del Hato National Astronomical Observatory of Venezuela,
operated by CIDA for the Ministerio de Ciencia y Tecnolog{\'\i}a,
and at the Fred Lawrence Whipple Observatory of the
Smithsonian Institution, USA.



\begin{thebibliography}{}

\bibitem[Alcal\'a et al.(1996)]{alcala96} Alcal\'a, J. M., Terranegra, L., Wichmann, R., Chavarr{\'\i}a-K., C., Krautter, et al. 1996, \aaps, 119, 7
\bibitem[Baade \& Minkowski(1937)]{baade37} Baade, W. \& Minkowski, R. 1937, \apj, 86, 119
\bibitem[Bally et al.(1987)]{bally87} Bally, J., Stark, A.~A., Wilson, R.~W., \& Langer, W.~D.\ 1987, \apjl, 312, L45
\bibitem[Baltay et al.(2002)]{baltay02} Baltay, C., Snyder, J.A., Andrews, P. et al. 2002, \pasp, 114, 780
\bibitem[Baraffe et al.(1998)]{baraffe98} Baraffe, I., Chabrier, G., Allard, F., \& Hauschildt, P.H. 1998, \aap, 337, 403
\bibitem[Barrado y Navascu\'es et al.(2003)]{barrado03} Barrado y Navascu\'es, D., B\'ejar, V. J. S., Mundt, R., Mart{\'\i}n, E. L., Rebolo, R., et al. 2003, \aap, 404, 171
\bibitem[Barrado y Navascu{\'e}s
\& Mart{\'{\i}}n(2003)]{barradomartin03} Barrado y Navascu{\'e}s, D., \& Mart{\'{\i}}n, E.~L.\ 2003, \aj, 126, 2997
\bibitem[Barrado y Navascu\'es et al.(2004)]{barrado04}  Barrado y Navascu\'es D., Stauffer J.~R., Bouvier J., Jayawardhana R., \& Cuillandre J-C. 2004, \apj, 610, 1064
\bibitem[Basri et al.(1991)]{basri91} Basri, G., Mart{\'\i}n, E. L., \& Bertout, C. 1991, \aap, 252, 625
\bibitem[B\'ejar et al.(1999)]{bejar99} B\'ejar, V. J. S., Zapatero Osorio, M. R., \& Rebolo, R. 1999, \apj, 521, 671
\bibitem[B{\'e}jar et al.(2003)]{bejar03} B{\'e}jar, V.~J.~S., Rebolo, R., Zapatero Osorio, M.~R., \& Caballero, J.~A.\ 2003, in {\sl The Future of Cool-Star Astrophysics: 12th Cambridge Workshop on Cool Stars, Stellar Systems, and the Sun}, eds.~A.~Brown, G.M.~Harper, \& T.R.~Ayres, (University of Colorado), 651
\bibitem[Blaauw(1964)]{blaauw64} Blaauw, A. 1964, \araa, 2, 213
\bibitem[Blaauw(1991)]{blaauw91} Blaauw, A. 1991, in {\it The Physics of Star Formation and Early Stellar Evolution}, eds. C. Lada and N.D. Kylafis, (Dordrecht: Kluwer), p. 125
\bibitem[Brice\~no et al.(1997)]{briceno97} Brice\~no, C., Hartmann, L., Stauffer, J., Gagne, M., Caillault, J.-P., \& Stern, A. 1997, \aj, 113, 740.
\bibitem[Brice\~no et al.(1999)]{briceno99} Brice\~no C., Hartmann L., Calvet N., \& Kenyon, S.  1999, \aj, 118, 1354
\bibitem[Brice\~no et al.(2001)]{briceno01} Brice\~no C., Vivas A.~K., Calvet N., Hartmann L., Pacheco, R., et al. 2001, \science, 291, 93
\bibitem[Brice{\~n}o et al.(2002)]{briceno02} Brice{\~n}o, C.,
Luhman, K.~L., Hartmann, L., Stauffer, J.~R.,
\& Kirkpatrick, J.~D.\ 2002, \apj, 580, 317
\bibitem[Brice\~no et al.(2005)]{briceno05} Brice\~no C., Calvet N., Hern\'andez J., Vivas A.~K., Hartmann L., et al. 2005, \aj, 129, 907
\bibitem[Brice\~no et al.(2007a)]{briceno07a} Brice\~no, C., Preibisch, T., Sherry, W.~H., Mamajek, E.~E., Mathieu, R.~D., Walter, F.~M., \& Zinnecker, H. 2007a, in {\it Protostars \& Planets V}, eds. B. Reipurth, D. Jewitt, \& K. Keil, (Tucson: University of Arizona Press), p. 345
\bibitem[Brice\~no et al.(2007b)]{briceno07b} Brice\~no C., Hartmann, L., Hern\'andez J., Calvet, N., Vivas A.~K., et al. 2007b, \apj, 661, 1119
\bibitem[Brice\~no et al.(2008)]{briceno08} Brice\~no, C. et al. 2008, AJ, in preparation
\bibitem[Brown et al.(1994)]{brown94} Brown, A.~G.~A., de Geus, E.~J., \& de Zeeuw, P.~T.\ 1994, \aap, 289, 101
\bibitem[Brown et al.(1999)]{brown99} Brown, A.~G., Walter, F.~M., \& Blaauw, A. 1999, in ASP Conf. Ser., {\it The Orion Complex Revisited}, ed. M.J. McCaughrean \& A. Burkert, unpublished
\bibitem[Caballero \& Solano(2008)]{caballero08} Caballero, J.~A. \& Solano, E.\ 2008, \aap, 485, 931
\bibitem[Calvet et al.(2002)]{calvet02} Calvet N., D'Alessio P., Hartmann L., Wilner D., Walsh A., \& Sitko M. 2002, \apj, 568, 1008
\bibitem[Calvet et al.(2005a)]{calvet05a} Calvet N., Brice\~no C., Hern\'andez J., Hoyer S., Hartmann L., et al. 2005a, \aj, 129, 935
\bibitem[Calvet et al.(2005b)]{calvet05b} Calvet N., D'Alessio P., Watson D.~M., Franco-Hern\'andez R., Furlan E.  et al. 2005b, \apj, 630, L185
\bibitem[Comer\'on et al.(1996)]{comeron96} Comeron, F., Rieke, G. H., \& Rieke, M. J. 1996, \apj, 473, 294
\bibitem[Crawford \& Barnes(1966)]{cb66} Crawford, D. L. \& Barnes, J. V. 1966, \aj, 71, 610
\bibitem[Dahari \& Lada(1999)]{dahari99} Dahari, D. B. \& Lada, E. A. 1999, American Astronomical Society, 195th AAS Meeting, \#79.13; Bulletin of the American Astronomical Society, 31, 1491
\bibitem[D'Alessio et al.(2005)]{dalessio05}  D'Alessio P., Hartmann L., Calvet N., Franco-Hern\'andez R., Forrest W. et al. 2005, \apj, 621, 461
\bibitem[Dolan \& Mathieu(1999)]{dom99} Dolan, C. J. \& Mathieu, R., D. 1999, \aj, 118, 2409
\bibitem[Dolan \& Mathieu(2001)]{dom01} --- 2001, \aj, 121, 2124
\bibitem[Dolan \& Mathieu(2002)]{dom02} --- 2002, \aj, 123, 387
\bibitem[Downes \& Keyes(1988)]{downes88} Downes, R. A. \& Keyes, C. D. 1988, \aj, 96, 777
\bibitem[Downes et al.(2006)]{downes2006} Downes, J.~J., Brice{\~n}o, C., \& Hern{\'a}ndez, J.\ 2006, Revista Mexicana de Astronomia y Astrofisica, Conference Series, 26, 37
\bibitem[Downes et al.(2008)]{downes2008} Downes, J.~J., Brice\~no, C., Hern\'andez, J., Calvet, N., Hartmann, L., \& Ponsot Balaguer, E.~A. 2008, \aj, 136, 51
\bibitem[Espaillat et al.(2007)]{espaillat07} Espaillat, C.,
Calvet, N., D'Alessio, P., Hern{\'a}ndez, J., Qi, C., et al. 2007, \apjl, 670, L135
\bibitem[Feigelson \& DeCampli(1981)]{feig89} Feigelson E.~D. \& DeCampli W.~M. 1981, \apj, 243, L89
\bibitem[Feigelson et al.(2002)]{feigelson02} Feigelson E.D. , Broos P., Gaffney III J.A., Garmire G., Hillenbrand L.A., et al. 2002, \apj, 574, 258
\bibitem[Feigelson et al.(2003)]{feigelson03} Feigelson, E. D., Gaffney, J. A., III, Garmire, G., Hillenbrand, L. A., \& Townsley, L. 2003, \apj, 584, 911
\bibitem[Freyberg \& Schmitt(1995)]{frey95} Freyberg, M. J. \& Schmitt, J. H. M. M. 1995, \aap, 296, 21
\bibitem[Gagn\'e \& Caillault(1994)]{gagne94} Gagn\'e M. \& Caillault J.-P. 1994, \apj, 437, 361
\bibitem[Garc{\'\i}a-L\'opez et al. (1994)]{garcialopez94} Garc{\'\i}a-L\'opez, R. J., Rebolo, R., \& Mart{\'\i}n, E. L. 1994, \aap, 282, 518
\bibitem[Gaustad et al.(2001)]{gaustad01} Gaustad, J.~E.,
McCullough, P.~R., Rosing, W., \& Van Buren, D.\ 2001, \pasp, 113, 1326
\bibitem[Genzel et al.(1981)]{grm81} Genzel, R., Reid, M. J., Moran, J.M., \& Downes, D. 1981, \apj, 244, 884
\bibitem[Genzel \& Stutzki(1989)]{ges89} Genzel, R. \& Stutzki, J. 1989, \araa, 27, 41
\bibitem[Getman et al.(2005)]{getman05} Getman K.V. , Feigelson E.D., Grosso N., McCaughrean M.J., Micela G., et al. 2005, \apjs, 160, 353
\bibitem[G\'omez \& Lada(1998)]{gomezlada98} G\'omez, M. \& Lada, C. J. 1998, \aj, 115, 1524
\bibitem[Gullbring et al.(1998)]{gullbring98} Gullbring, E., Hartmann, L., Brice\~{n}o, C., \& Calvet, N., 1998, \apj, 492, 323
\bibitem[Gutermuth et al.(2006)]{gutermuth06} Gutermuth, R. A., Pipher, J. L., Myers, P. C., Megeath, T. S., Allen, L. E., \& Allen, T. 2006, AAS, 208, No.8.07
\bibitem[Haisch et al.(2000)]{haisch00} Haisch, K. E., Jr., Lada, E. A., \& Lada, C. J. 2000, \aj, 120, 1396
\bibitem[Haisch et al.(2001)]{haisch01} Haisch, K. E., Jr., Lada, E. A., Pi\~na, R. K., Telesco, C. M., \& Lada, C. J. 2001, \aj, 121, 1512
\bibitem[Hardie et al.(1964)]{hht64} Hardie, R. H., Heiser, A. M., \& Tolbert, C. R. 1964, \apj, 140, 1472
\bibitem[Hartmann et al.(1998)]{hartmann98} Hartmann, L., Calvet, N., Gullbring, E. \& D'Alessio, P, 1998, \aj, 495, 385
\bibitem[Hartmann et al.(2005)]{hartmann05} Hartmann, L., Calvet, N.,  Watson, D.M. et al. 2005, \apj, 628, L147
\bibitem[Harvey et al.(1979)]{harvey79} Harvey, P. M., Campbell, M. F., Hoffmann, W. F., Thronson, H. A., Jr., \& Gatley, I. 1979, \apj, 229, 990
\bibitem[Herbig(1962)]{herbig62} Herbig, G.H. 1962, Adv. Astron. Astrophys. 1, 47
\bibitem[Herbig \& Terndrup(1986)]{herbig86} Herbig, G. H. \& Terndrup, D.M. 1986, \apj, 307, 609
\bibitem[Herbig \& Bell(1988)]{herbig88} Herbig G.~H. \& Bell K.~R. 1988, Lick Observatory Bulletin No. 1111, Lick Observ., Santa Cruz
\bibitem[Herbig \& Kuhi(1963)]{herbig63} Herbig, G. H. \& Kuhi, L. V. 1963, \apj, 137, 398
\bibitem[Herbig \& Rao(1972)]{herbig72} Herbig, G.~H. \& Rao, N.~K. 1972, \apj, 174, 401
\bibitem[Herbst et al.(2000)]{herbst00} Herbst, W., Rhode, K. L., Hillenbrand, L. A., \& Curran, G. 2000, \aj, 119, 261
\bibitem[Hern\'andez et al.(2005)]{hernandez05} Hern\'andez J., Calvet, N., Hartmann, L., Brice\~no, C., Sicilia-Aguilar, A., \& Berlind, P. 2005, \aj, 129, 856
\bibitem[Hern{\'a}ndez et al.(2006)]{hernandez06} Hern{\'a}ndez,
J., Brice{\~n}o, C., Calvet, N., Hartmann, L., Muzerolle, J.,
\& Quintero, A.\ 2006, \apj, 652, 472
\bibitem[Hern{\'a}ndez et al.(2007a)]{hernandez07a} Hern{\'a}ndez, J., Hartmann, L., Megeath, T., Gutermuth, R., Muzerolle, J. et al. 2007a, \apj, 662, 1067
\bibitem[Hern\'andez et al.(2007b)]{hernandez07b} Hern\'andez, J., Calvet, N., Brice\~no, C.., Hartmann, L., Vivas, A. K., et al. 2007b, \apj, 671, 1784
\bibitem[Hillenbrand(1997)]{hill97} Hillenbrand L.~A. 1997, \aj, 113, 1733
\bibitem[Hillenbrand et al.(1998)]{hill98} Hillenbrand, L. A., Strom, S. E., Calvet, N., Merrill, K. M., Gatley, I. et al. 1998, \aj, 116, 1816
\bibitem[Jeffries et al.(2006)]{jeffries06} Jeffries, R. D., Maxted, P. F. L., Oliveira, J. M., \& Naylor, T. 2006, \mnras, 371, 6
\bibitem[Johnson(1965)]{johnson65} Johnson, H. M. 1965, \apj, 142, 964
\bibitem[Joy(1945)]{joy45} Joy A.~H. 1945, \apj, 102, 168
\bibitem[Kenyon \& Bromley(2005)]{kenyonbromley05} Kenyon, S.~J. \& Bromley, B.~C.\ 2005, \aj, 130, 269
\bibitem[Kenyon \& Hartmann(1995)]{kh95} Kenyon, S.  J. \& Hartmann, L., 1995, \apjs, 101, 117
\bibitem[Kenyon et al.(2005)]{keny05} Kenyon, M. J., Jeffries, R. D., Naylor, T., Oliveira, J. M., \& Maxted, P. F. L. 2005, \mnras, 356, 89
\bibitem[Kirkpatrick et al.(1999)]{kir99} Kirkpatrick, J. D., Reid, I. N., Liebert, J., Cutri, R. M., Nelson, B. et al. 1999, \apj, 519, 802
\bibitem[Kogure et al.(1989)]{kyw89} Kogure, T., Yoshida, S., Wiramihardja, S., Nakano, M., Iwata, T. \& Ogura, K. 1989, PASJ, 41, 1195
\bibitem[Kroupa(2002)]{kroupa02} Kroupa P. 2002, \science, 295, 82
\bibitem[Kutner et al.(1971)]{kut77} Kutner, M. L., Tucker, K. D., Chin, G., \& Thaddeus, P. 1977, \apj, 215, 521
\bibitem[Lada \& Lada(2003)]{ladalada03} Lada C.~J. \& Lada, E.~A. 2003, \araa, 41, 57
\bibitem[Lamm et al.(2004)]{lamm04} Lamm, M. H., Bailer-Jones, C. A. L., Mundt, R., Herbst, W., \& Scholz, A. 2004, \aap, 417, 557
\bibitem[Levine et al.(2006)]{levine06} Levine, J. L., Steinhauer, A., Elston, R. J., \& Lada, E. A. 2006, \apj, 646, 1215
\bibitem[Luhman et al.(2003a)]{luh03a} Luhman, K.~L.,
Brice{\~n}o, C., Stauffer, J.~R., Hartmann, L., Barrado y Navascu{\'e}s,
D., \& Caldwell, N. 2003a, \apj, 590, 348
\bibitem[Luhman et al.(2003b)]{luh03b} Luhman, K.~L., Stauffer,
J.~R., Muench, A.~A., Rieke, G.~H., Lada, E.~A., et al. 2003b, \apj, 593, 1093
\bibitem[Luhman(2004a)]{luh04a} Luhman, K. L. 2004a, \apj, 602, 816
\bibitem[Luhman(2004b)]{luh04b} Luhman, K. L. 2004b, \apj, 617, 1216
\bibitem[Luhman et al.(2006)]{luhman06} Luhman, K.~L., Whitney,
B.~A., Meade, M.~R., Babler, B.~L., Indebetouw, R., Bracker, S.,
\& Churchwell, E.~B.\ 2006, \apj, 647, 1180
\bibitem[Lynds(1962)]{lynds62} Lynds, B. T. 1962, \apjs, 7, 1
\bibitem[Maddalena et al.(1986)]{maddalena86} Maddalena, R. J., Morris, M., Moscowitz, J., \& Thaddeus, P. 1986, \apj, 303, 375
\bibitem[Maheswar et al.(2003)]{maheswar03} Maheswar, G., Manoj, P. \& Bhatt, H. C. 2003, \aap, 402, 963
\bibitem[Mart{\'\i}n et al.(1994)]{martin94} Mart{\'\i}n, E. L., Rebolo, R., Magazzu, A., \& Pavlenko, Ya. V. 1994, \aap, 282, 503
\bibitem[Mart{\'\i}n (1997)]{martin97} Mart{\'\i}n, E. L. 1997, \aap, 321, 492
\bibitem[McGehee et al.(2005)]{mcgehee05} McGehee P.~M., West A.~A., Smith J.~A., Anderson, K.~S.~J., \& Brinkmann J. 2005, \aj, 130, 1752
\bibitem[McGehee(2006)]{mcgehee06} McGehee, P.~M. 2006, \aj, 131, 2959
\bibitem[Muzerolle et al.(2000)]{muzerolle00} Muzerolle, J., Calvet, N., Brice\~{n}o, C., Hartmann, L., \& Hillenbrand, L. 2000, \apj, 535, L47
\bibitem[Muzerolle et al.(2005)]{muzerolle05} Muzerolle, J., Young, E., Megeath, S. T., \& Allen, L. 2005, in {\it Star Formation in the Era of Three Great Observatories}, meeting abstracts from the conference held July 13-15, 2005 in Cambridge, MA. http://cxc.harvard.edu/stars05/agenda/program.html,
p. 41.
\bibitem[Pollack et al.(1996)]{pollack96} Pollack, J.~B.,
Hubickyj, O., Bodenheimer, P., Lissauer, J.~J., Podolak, M.,
\& Greenzweig, Y.\ 1996, Icarus, 124, 62
\bibitem[Preibisch et al.(2005)]{preibisch05} Preibisch, T., McCaughrean, M. J., Grosso, N., Feigelson, E. D., Flaccomio, E., et al. 2005, \apjs, 160, 582
\bibitem[Ram{\'\i}rez et al.(2004)]{ramirez04} Ram{\'\i}rez S.V. , Rebull L., Stauffer J., Strom S., Hillenbrand L., et al. 2004, \apj, 128, 787
\bibitem[Randich et al.(1997)]{randich97} Randich, S., Aharpour, N., Pallavicini, R., Prosser, C. F., \& Stauffer, J. R. 1997, \aap, 323, 86
\bibitem[Rebull et al.(2000)]{rebull00} Rebull, L. M., Hillenbrand, L. A., Strom, S. E., Duncan, D. K., Patten, B. M., et al. 2000, \apj, 119, 3026
\bibitem[Sanduleak(1971)]{san71} Sanduleak N. 1971, \pasp, 83, 95
\bibitem[Schlegel et al.(1998)]{schlegel98} Schlegel, D.J., Finkbeiner, D.P., \& Davis, M. 1998, \apj, 500, 525
\bibitem[Scholz \& Eisl{\"o}ffel(2005)]{scholz05} Scholz, A. \& Eisl{\"o}ffel, J.\ 2005, \aap, 429, 1007
\bibitem[Sherry(2003)]{sherry03} Sherry W.~H. 2003, PhD Thesis, SUNY, Stony Brook
\bibitem[Sherry et al.(2004)]{sherry04} Sherry, W. H., Walter, F. M., \& Wolk, S. J. 2004, \aj, 128, 2316
\bibitem[Sicilia-Aguilar et a.(2005)]{aurora05} Sicilia-Aguilar, A., Hartmann, L. W., Szentgyorgyi, A. H., Fabricant, D. G., Fur\'esz, G., et al. 2005, \aj, 129, 363
\bibitem[Siess et al.(2000)]{siess00}  Siess, L., Dufour, E., \& Forestini, M.\ 2000, \aap, 358, 593
\bibitem[Skinner et al.(2003)]{skinner03} Skinner, S. Gagn\'e, M., \& Belzer, E. 2003, \apj, 598, 375
\bibitem[Slesnick et al.(2006)]{slesnick06} Slesnick, C. L., Carpenter, J. M., \& Hillenbrand, L. A. 2006, \aj, 131, 3016
\bibitem[Soderblom et al.(1993)]{soderblom93} Soderblom, D. R., Jones, B. F., Balachandran, S., Stauffer, J. R., Duncan, D. K., et al. 1993, \aj, 106, 1059
\bibitem[Stassun et al.(2006)]{stassun06} Stassun, K. G., van den Berg, M., Feigelson, E., \& Flaccomio, E. 2006, \apj, 649, 914
\bibitem[Stauffer et al. (1997)]{stauffer97} Stauffer, J. R., Hartmann, L. W., Prosser, C. F., Randich, S., Balachandran, S., et al., 1997, \apj, 479, 776
\bibitem[Stauffer et al. (1998)]{stauffer98} Stauffer, J. R., Schultz, G., \& Kirkpatrick, J. D. 1998, \apj, 499, 199
\bibitem[Stephenson(1986)]{stephenson86} Stephenson, C. B. 1986, \apj, 300, 779.
\bibitem[Sterzik et al.(1995)]{sterzik95} Sterzik, M. F., Alcal\'a, J. M.,
Neuh\"auser, R., \& Schmitt, J. H. M. M. 1995, \aap, 297, 418
\bibitem[Tsujimoto et al. (2002)]{tsujimoto02} Tsujimoto M. , Koyama K., Tsuboi Y., Goto M., \& Kobayashi N. 2002, \apj, 566, 974
\bibitem[Uchida et al.(2004)]{uchida04}  Uchida K.~I., Calvet N., Hartmann L., Kemper F., Forrest W.~J., et al. 2004, \apjs, 154, 439
\bibitem[Voges et al.(1999)]{voges99} Voges W. , Aschenbach B., Boller T., Braeuninger H., Briel U., et al. 1999, \aap, 349, 389
\bibitem[Walker(1969)]{walker69}  Walker, M. F. 1969 \apj, 155, 447
\bibitem[Walter \& Kuhi(1981)]{wku81} Walter F.~M. \& Kuhi L.~V. 1981, \apj, 250, 254
\bibitem[Walter \& Myers(1986)]{waltmyers86} Walter F.~M. \& Myers P.~C. 1986, in {\it IV Cambridge Workshop on Cool Stars, Stellar Systems, and the Sun}, M. Zeilik \& D.~M. Gibson (eds.), 254, p. 55. Springer-Verlag, Berlin-Heidelberg-New York.
\bibitem[Walter et al.(1997)]{walter97} Walter, F. M., Wolk, S. J., Freyberg, M., \& Schmitt, J. H. M. M. 1997, Memorie della Societa Astronomia Italiana, 68, 1081
\bibitem[Warren \& Hesser(1977)]{wh77} Warren, W.H. \& Hesser, J.E. 1977, \apjs, 34, 115
\bibitem[Warren \& Hesser(1978)]{wh78} Warren, W.H. \& Hesser, J.E. 1978, \apjs, 36, 497
\bibitem[Weaver \& Babcock(2004)]{weav04} Weaver W.~B. \& Babcock A. 2004, \pasp, 116, 1035
\bibitem[Weidenschilling(1997)]{weid97} Weidenschilling,
S.~J.\ 1997, Lunar and Planetary Institute Conference Abstracts, 28, 1517
\bibitem[White \& Basri(2003)]{wba03} White, R.J. \& Basri, G. 2003, \apj, 582, 1109
\bibitem[Wiramihardja et al.(1989)]{wky89} Wiramihardja, S., Kogure,
T., Yoshida, S., Ogura, K., \& Nakano, M. 1989, in IAUS No. 135{\it
Interstellar Dust}, NASA N91-14897 06-88, p. 239
\bibitem[Wiramihardja et al. (1991)]{wky91} Wiramihardja, S., Kogure, T., Yoshida, S., Nakano, M., Ogura, K., \& Iwata, T. 1991, PASJ, 43, 27
\bibitem[Wiramihardja et al.(1993)]{wky93} Wiramihardja, S., Kogure, T., Yoshida, S., Ogura, K., \& Nakano, M. 1993, PASJ, 45, 643
\bibitem[Wolk et al.(2005)]{wolk05} Wolk, S. J., Harnden, F. R., Jr., Flaccomio, E., Micela, G., Favata, F., et al. 2005, \apjs, 160, 423
\bibitem[Yamauchi et al.(1996)]{yamauchi96} Yamauchi S. , Koyama K., Sakano M., \& Okada K. 1996, \pasj, 48, 719
\bibitem[Zapatero Osorio et al.(2002)]{zapatero02} Zapatero Osorio, M. R., B\'ejar, V. J. S., Pavlenko, Ya., Rebolo, R., Allende Prieto, C., et al. 2002, \aap, 384, 937

\end{thebibliography}
\end{document}